\documentclass[12pt]{article}

\pdfoutput=1
\usepackage{color}
\usepackage{amsmath, amssymb}
\usepackage{epsfig, palatino}
\usepackage{pstricks,pst-node,pst-tree}
\usepackage{epic}

\usepackage{mathrsfs}

\usepackage{ae}
\usepackage[T1]{fontenc}
\usepackage[ansinew]{inputenc}
\usepackage{amsmath}
\usepackage{amssymb}
\usepackage{graphicx}
\usepackage{color}
\definecolor{darkblue}{cmyk}{0.9,0.9,0,0}
\usepackage[colorlinks=true,linkcolor=darkblue,citecolor=darkblue,urlcolor=darkblue]{hyperref}
\usepackage{cite}
\usepackage{hyperref}
\usepackage{wasysym}

\newcommand{\comment}[1]{}

\newcommand{\beq}{\begin{equation}}
\newcommand{\eeq}{\end{equation}}
\newcommand{\beqq}{\begin{equation*}}
\newcommand{\eeqq}{\end{equation*}}
\newcommand\beqa{\begin{eqnarray}}
\newcommand\eeqa{\end{eqnarray}}
\newcommand\beqaa{\begin{eqnarray*}}
\newcommand\eeqaa{\end{eqnarray*}}
\newcommand\bea{\begin{array}}
\newcommand\eea{\end{array}}

\def\XXint#1#2#3{{\setbox0=\hbox{$#1{#2#3}{\int}$}
\vcenter{\hbox{$#2#3$}}\kern-.5\wd0}}

\newcommand{\neqa}{\nonumber\end{eqnarray}}

\renewcommand{\d}{\partial}

\newcommand{\<}{{\langle}}
\renewcommand{\>}{{\rangle}}

\newcommand{\re}{\relax{\rm I\kern-.18em R}}

\renewcommand{\sp}{p\hspace{-.40em}/}

\def\su2{{SU(2)}}

\def\[{\left[}
\def\]{\right]}

\def\({\left(}
\def\){\right)}
\def\[{\left[}
\def\]{\right]}

\def\<{\langle}
\def\>{\rangle}

\def\i2{\frac{i}{2}}

\def\spi{\relax{\rm \pi\kern-0.5em /}}
\def\sA{\relax{\rm A\kern-0.5em /}}
\def\sp{\relax{\rm p\kern-0.5em /}}
\def\sd{\relax{\rm \d\kern-0.5em /}}
\def\sk{\relax{\rm k\kern-0.5em /}}
\def\sn{\relax{\rm n\kern-0.5em /}}
\def\sl{\relax{\rm l\kern-0.5em /}}
\def\sP{\relax{\rm P\kern-0.7em /}}
\def\sBethe{\relax{\rm \Bethe\kern-0.5em /}}

\def\2F1{\,_2{\rm F}_1}

\newcommand{\rd}{\mathrm{d}}

\newcommand{\rE}{\mathrm{E}}

\newcommand{\rJ}{\mathrm{J}}

\newcommand{\rT}{\mathrm{T}}

\newcommand\spa{,\qquad }
\newcommand{\der}{\mathrm{d}}

\newcommand{\group}[2]{\mathrm{#1}(#2)}

\usepackage{varioref}
\usepackage{makeidx}
\makeindex

\usepackage[english]{babel}
\begin{document}

\thispagestyle{empty}

\renewcommand{\thefootnote}{\fnsymbol{footnote}}
\setcounter{page}{1}
\setcounter{footnote}{0}
\setcounter{figure}{0}
\begin{flushright}
\textcolor{gray}{CERN-TH-2020-221}
\end{flushright}
\begin{center}
$$$$
{\Large\textbf{\mathversion{bold}
Geometrizing non-relativistic bilinear deformations}\par}

\vspace{1.0cm}

\textrm{Dennis Hansen$^{\mathcal{E}^{\mu}}$\quad Yunfeng Jiang$^{\mathcal{P}^{1},\mathcal{P}^{2}}$\quad Jiuci Xu$^{\mathcal{J}^{1},\mathcal{J}^{2}}$}
\\ \vspace{1.2cm}
\footnotesize{
\textit{$^{\mathcal{E}^{\mu}}$
Institut f\"ur Theoretische Physik, Eidgen\"ossische Technische Hochschule Z\"urich,\\
Wolfgang-Pauli-Strasse 27, 8093 Z\"urich, Switzerland\\
\vspace{4mm}
$^{\mathcal{P}^{1}}$Department of Theoretical Physics, CERN,\\
1 Esplanade des Particules, Geneva 23, CH-1211, Switzerland\\
$^{\mathcal{P}^{2}}$ Shing-Tung Yau Center and School of physics, Southeast University, Nanjing 210096, China\\
\vspace{4mm}
$^{\mathcal{J}^{1}}$ Department of Physics, University of California, Santa Barbara, CA 93106, USA\\
$^{\mathcal{J}^{2}}$ University of Science and Technology of China, 96 Jinzhai Road
230026, Hefei, Anhui, China
\vspace{4mm}
}

\par\vspace{1.5cm}

\textbf{Abstract}\vspace{2mm}
}\end{center}
We define three fundamental solvable bilinear deformations for any massive non-relativistic 2d quantum field theory (QFT). They include the $\mathrm{T}\overline{\mathrm{T}}$ deformation and the recently introduced hard rod deformation. We show that all three deformations can be interpreted as coupling the non-relativistic QFT to a specific Newton--Cartan geometry, similar to the Jackiw--Teitelboim-like gravity in the relativistic case. Using the gravity formulations, we derive closed-form deformed classical Lagrangians of the Schr\"odinger model with a generic potential. We also extend the dynamical change of coordinate interpretation to the non-relativistic case for all three deformations. The dynamical coordinates are then used to derive the deformed classical Lagrangians and deformed quantum S-matrices.

\noindent

\setcounter{page}{1}
\renewcommand{\thefootnote}{\arabic{footnote}}
\setcounter{footnote}{0}

 \def\nref#1{{(\ref{#1})}}

\newpage

\tableofcontents

    \parskip 5pt plus 1pt   \jot = 1.5ex
\section{Introduction}
Recent studies of solvable irrelevant deformations of relativistic quantum field theories have extended our understanding of them.
The most studied example of such deformations is the $\rT\overline{\rT}$ deformation \cite{Smirnov:2016lqw,Cavaglia:2016oda}, which can be defined for any Lorentz invariant 2d QFT with a local stress energy tensor. For theories with additional symmetries, similar deformations such as the $J\bar{T}$ \cite{Guica:2017lia}, $JT_a$ deformations \cite{Anous:2019osb} and the ones constructed by higher conserved currents \cite{LeFloch:2019rut,Conti:2019dxg} have been studied. All these deformations share similar features. They all modify the UV behavior of the QFTs and lead to non-local theories, yet they are under good analytical control due to the deformations' solvability.

Surprisingly, such deformations can be defined not only for 2d relativistic QFTs but also for a much broader class of theories. These include non-Lorentz invariant QFTs \cite{Cardy:2018jho,Cardy:2020olv}, non-relativistic quantum many body systems such as the Bose gas \cite{Jiang:2020nnb} and lattice models like quantum spin chains \cite{Pozsgay:2019ekd,Marchetto:2019yyt}. Furthermore, it was recently shown \cite{Jiang:2020nnb,Cardy:2020olv} that the deformed 1d Bose gas share many qualitative features with $\rT\overline{\rT}$ deformed relativistic QFTs, such as the break down of UV physics for the spectrum and the Hagedorn behavior for thermodynamics. These findings hint that what we have seen so far from solvable deformations of relativistic QFTs is only the tip of the iceberg - the structure and main features of such deformations can be extended to a much wider setting.\par

One of the important lessons we learned from relativistic QFT is that $\rT\overline{\rT}$ deformation can be reformulated as coupling the QFT to certain special 2d topological gravity theory, both in flat \cite{Dubovsky:2017cnj,Dubovsky:2018bmo} and curved space \cite{Tolley:2019nmm,Caputa:2020lpa}, at least classically. Similar interpretations also holds for $JT_a$ deformation \cite{Anous:2019osb} where the gravity theory also involves an additional $\mathrm{U}(1)$ gauge field. The 2d gravity formulation is tightly related to the random geometry interpretation of $\rT\overline{\rT}$ deformation \cite{Cardy:2018sdv} and the dynamical change of coordinates \cite{Conti:2018tca,Guica:2019nzm,Cardy:2019qao,Caputa:2020lpa}. These formulations offer us a more geometrical understanding of the $\rT\overline{\rT}$ deformation.\par

It is natural to ask whether similar geometrical interpretations exist for other settings. This paper is the first step towards such an interesting goal by giving an affirmative answer in the context of non-relativistic QFTs. It is probably not too surprising that geometrical interpretations exist for non-Lorentz invariant QFTs, as the random geometry interpretation was already pointed out in \cite{Cardy:2018jho}. Nevertheless, notice that the random geometry picture follows directly from a Hubbard--Stratonovich transformation of the $\rT\overline{\rT}$ operator.
It is not at all obvious from this what the gravity theory the QFT couples to is. Besides, given that the seed and deformed theories are not Lorentz invariant, the gravity theory cannot be a usual Einstein--Hilbert type gravity.
A more natural candidate, in this case, is Newton--Cartan geometry, which is manifestly covariant under non-relativistic symmetry \cite{Cartan1,Cartan2,Andringa:2010it,Christensen:2013lma,Hansen:2018ofj}.\par

We will show that for non-relativistic QFTs, an elegant gravity interpretation for $\rT\bar{\rT}$- and two other solvable deformations to be defined shortly indeed exists in the framework of Type I Newton--Cartan geometry. For the $\rT\overline{\rT}$ deformation, the gravity theory written in the first-order formalism takes the same form as its relativistic counterpart \cite{Dubovsky:2018bmo,Tolley:2019nmm}. The gravity theory for the other two deformations is similar to the ones for $JT_a$ deformed relativistic QFTs \cite{Anous:2019osb}. From the gravity formulation, the dynamical change of coordinates interpretation naturally follows \cite{Tolley:2019nmm,Caputa:2020lpa}. The dynamical coordinates provide a powerful tool to compute several important quantities, such as the deformed classical Lagrangian and quantum S-matrices. \par

Apart from extending what we have learned from relativistic QFTs to the non-relativistic settings, the gravity formulation for the non-relativistic QFTs is also interesting in its own right. To start with, most field theories one encounters in condensed matter physics are not Lorentz invariant. $\rT\overline{\rT}$ deformations for Galilean or Bargmann (the central extension of the Galilei group \cite{Bargmann:1954gh}) invariant QFTs are definitely interesting to study. In particular, such QFTs are closely related to condensed matter systems that can be realized in experiments, one may gain more physical intuitions about $\rT\overline{\rT}$ deformation by studying such systems. Notice that Newton--Cartan gravity has already played a role in condensed matter systems such as quantum Hall effect \cite{Son:2013rqa} and unitary Fermi gas \cite{Son:2005rv}. It would be fascinating to make connections to these fields. \par

From the gravity point of view, Newton--Cartan geometry is different from Einstein--Hilbert gravity and exhibits new features.
One important difference is that there is a built-in $\mathrm{U}(1)$ symmetry in massive non-relativistic QFTs, which corresponds to the conservation of mass or particle number.
Correspondingly, an $\mathrm{U}(1)$ gauge field is an essential ingredient in so-called Type I Newton--Cartan geometry, which displays local Bargmann symmetry.
Therefore, any Bargmann invariant 2d QFTs have at least three fundamental local symmetry currents, which correspond to mass, momentum, and energy conservation. This implies,
apart from the $\rT\overline{\rT}$ deformation, we can always define two other solvable bilinear deformations constructed from the mass current with the momentum or energy currents. We shall call these two deformations the hard rod deformation and the JE deformation, respectively. As we alluded before, these two deformations also allow a gravity interpretation in Newton--Cartan geometry. We study these three fundamental deformations in parallel. The hard rod and JE deformations are new in non-relativistic QFTs. They are similar to the $JT_a$ deformation of relativistic QFTs. The difference is that in the non-relativistic setting such $\mathrm{U}(1)$ symmetry is built-in and does not require further assumptions.\par

Interestingly, the hard rod deformation was constructed very recently in \cite{Cardy:2020olv,Jiang:2020nnb}. It is found that this deformation has the effect of deforming the point particles to finite size hard rods\footnote{To be more precise, this is true for one sign of the deformation parameter. For the other sign, the hard rod deformation increases the distance between particles.}. We will confirm this intuition and show that it can be formulated equivalently as coupling the undeformed theory to a specific Newton--Cartan geometry. Intuitively, this relation can be understood as follows. Traditionally, the hard rod model can be solved by performing a change of coordinates which eliminates the sizes of the rods and leaves only the free space (see for example \cite{hardrod}). In the new coordinate, the rods become point particles and the model can be solved readily. Therefore, the crucial point here is the change of coordinates. Since this change amounts to a redefinition of the length, it can be formulated as putting the theory on a different geometry where the metric is defined differently. A covariant way to formulate this intuition is precisely coupling the original theory to the proper Newton-Cartan geometry. We will see that the change of coordinate we mentioned above appear naturally from the 2d gravity formulation in section~\ref{sec:DCC}.\par

The $\rT\overline{\rT}$ deformation can be understood in the same way. Under $\rT\overline{\rT}$ deformation, point particles also become hard rods\footnote{Again for the sign where the effective length of the system becomes smaller. For the other sign, the space between the particles is increased.}. The only difference from the hard rod deformation is that now the size of each rod is no longer a fixed number, but depends on the energy of the particle (or the rod), which needs to be determined self-consistently. Therefore we obtain certain `dynamical' hard rod model. This new kind of hard rod model can also be solved by a change of coordinate, but now the new coordinates depend on the stress-energy tensor of the theory and thus become field-dependent, or dynamical. This gives a clear physical understanding of the dynamical change of coordinates, both in the non-relativistic case and the Lorentz invariant case. This also explains why the $\rT\overline{\rT}$ deformed theory is non-local. It is simply because we are describing finite size objects such as strings or hard rods in the deformed theory.

The paper is structured as follows: In section~\ref{sec:NC_2d}, we give a pedagogical review of Type I Newton--Cartan gravity in 2d.
In section~\ref{sec:bilinear}, we derive the gravity formulations for the three bilinear deformations. This is achieved by viewing these deformations' definitions as equations for the classical action and solving them formally by a heat-kernel-like approach. In section~\ref{sec:defLag}, we apply and test our gravity formulation by deriving closed-form classical Lagrangians. We compare the results with the ones obtained from a direct approach and find a perfect match. In section~\ref{sec:DCC}, we derive the dynamical change of coordinates from the gravity formulation. The dynamical coordinates provide yet another approach to derive the deformed Lagrangian. In addition, we derive the deformed quantum S-matrices using the dynamical coordinates. We conclude in section~\ref{sec:concl} and discuss future directions. Appendix \ref{sec:Newton-Cartan_app} is dedicated to a more detailed introduction to Newton--Cartan geometry.

\section{Newton--Cartan gravity in two dimensions}
\label{sec:NC_2d}
We give a pedagogical and minimal review on Newton--Cartan geometry suitable for our aims in the subsequent sections.
Only the formalism required for that is introduced.
A general review can be found in Appendix \ref{sec:Newton-Cartan_app}.

\subsection{Geometric content}
Newton--Cartan (NC) geometry is the natural framework to study non-relativistic field theories and their coupling to gravity.
It is a covariant formulation where the Galilean group is a subgroup of the local symmetry group.
It was first obtained as a geometrization of Newtonian spacetime by Cartan \cite{Cartan1,Cartan2}, but has seen a revival in recent years.
There is extensive literature on the subject and its applications to a wide selection of areas such as generalized holography \cite{Son:2008ye,Taylor:2008tg,Herzog:2008wg,Balasubramanian:2008dm,Alishahiha:2009np,Janiszewski:2012nb,Griffin:2012qx,Christensen:2013lma,Christensen:2013rfa,Hartong:2014pma,Taylor:2015glc,Harmark:2018cdl}, condensed matter theory \cite{Son:2013rqa,Jensen:2014aia,Hartong:2014oma,Geracie:2014nka,Gromov:2014vla,Laurila:2020yll}, hydrodynamics \cite{Jensen:2014ama,Geracie:2015xfa,Hartong:2016nyx,deBoer:2017ing,deBoer:2017abi,Armas:2019gnb,deBoer:2020xlc}, string theory \cite{Gomis:2000bd,Andringa:2012uz, Bagchi:2013bga,Harmark:2017rpg,Bergshoeff:2018yvt,Blair:2019qwi,Gallegos:2019icg,Gomis:2019zyu,Roychowdhury:2020kma} and more \cite{Morand:2017fnv,Cho:2019ofr,Berman:2019izh,Grumiller:2020elf,Gomis:2020wxp}.
We shall consider the 2-dimensional gravity relevant to our aims.

In 2d NC geometry, there are two fundamental tensors:
\begin{itemize}
    \item the clock-form $\tau_\mu$, giving the local flow of time and
    \item the spatial vector $e^\mu$, giving the local space direction.
\end{itemize}
Newton--Cartan geometry is defined by requiring that they satisfy the fundamental orthogonality relation
\begin{equation} \label{eq:NC_ortho_def}
    \tau_\mu e^\mu = 0.
\end{equation}
This implies that there is indeed a preferred direction of time in NC geometry, and the structure is implemented covariantly.
However, we may only define projective inverses $-v^\mu, e_\mu$ that preserves \eqref{eq:NC_ortho_def}.
The ambiguity in selecting the inverses is exactly the Galilean boost freedom, which is also referred to as Milne boosts in the literature \cite{Jensen:2014aia}.

When grouped together, the fields naturally form Galilean zweibeine defined as
\begin{align}
E_{\mu}^{A} & =  \left(\tau_{\mu},\,e_{\mu}\right), \qquad
E_{A}^{\mu} = \left(-v^{\mu},\,e^{\mu}\right),
\end{align}
and satisfying the completeness relations
\begin{equation} \label{eq:NC_comleteness_rel0}
    E_{\mu}^{A} E^{\mu}_{B} = \delta^A_B,\qquad
    E_{\mu}^{A} E^{\nu}_{A} = \delta_\mu^\nu .
\end{equation}
Written out in components these are
\begin{align} \label{eq:NC_comleteness_rel1}
    &\tau_\mu e^\mu = 0, \qquad
    \tau_\mu v^\mu = -1, \qquad
    e_\mu e^\mu = 1, \qquad
    e_\mu v^\mu = 0, \qquad
    -\tau_\mu v^\nu + e_\mu e^\nu = \delta_\mu ^\nu.
\end{align}
The zweibeine transform under diffeomorphisms and local Galilean transformations as
\begin{align}
\delta\tau_{\mu} & =  \mathcal{L}_{\xi}\tau_{\mu}, \qquad
\delta e_{\mu}  =  \mathcal{L}_{\xi}e_{\mu} + \lambda\tau_{\mu}, \label{eq:trafo_NC_fields_1}\\
\delta v^{\mu}  &= \mathcal{L}_{\xi}v^{\mu} + \lambda e^{\mu},\qquad
\delta e^{\mu}  =  \mathcal{L}_{\xi}e^{\mu},\label{eq:trafo_NC_fields_4}
\end{align}
where $\delta x^\mu := \xi^\mu$ is a vector field so that $\mathcal{L}_\xi$ is the Lie derivative
\begin{equation}\label{eq:Lie_derivative_diffeo}
    \mathcal{L}_{\xi} X^\mu{}_\nu = \xi^\rho \partial_\rho X^\mu{}_\nu - \partial_\rho\xi^\mu  X^\rho{}_\nu + \partial_\nu\xi^\rho  X^\mu{}_\rho,
\end{equation}
that generates infinitesimal diffeomorphisms of the zweibeine and $\lambda=\lambda(x^0,x^1)$ is a local Galilean boost.
Contrary to the relativistic case, the boost does not affect all the components in the same way --- the hallmark of non-relativistic physics.
We see that $\tau_{\mu},\, e^{\mu}$ indeed transform as tensors since they are invariant under local Galilean transformations.
The general dimensional case where spatial rotations also enter can be found in Appendix \ref{sec:NC_geometries_app}.

A generalization of the Newtonian potential $\Phi$ also arises.
The principle of covariance tells us that it must be obtained as a projection of a gauge field.
Indeed one finds that it is given by
\begin{equation}
    \Phi=-v^{\mu}m_{\mu},
\end{equation}
where $m_\mu$ is a $\mathrm{U}(1)$ gauge field known as the mass (or particle number) gauge field.
$m_\mu$ transforms as
\begin{equation}
    \delta m_{\mu}  =  \mathcal{L}_{\xi}m_{\mu}+\lambda e_{\mu}+\partial_{\mu}\sigma,
\end{equation}
where $\sigma$ is a $\mathrm{U}(1)$ transformation parameter.
From a more group theoretical perspective, $m_\mu$ is the gauge connection associated with the central charge of the Bargmann group.
The Bargmann group is the non-trivial central extension of the Galilean group, and the extra generator corresponds to mass or particle number conservation \cite{Bargmann:1954gh}.
We review these groups in Appendix \ref{sec:NR_groups}.

In other words, the geometry we have described above is the result of gauging the Bargmann group, a useful approach that has been studied in for example  \cite{Duval:1984cj,Andringa:2010it, Hartong:2015zia}.
This also makes it clear that $m_\mu$ is an integral part of the geometry.

\subsection{Flat Newton--Cartan spacetime}\label{sec:NC_flat_spacetime}
An important special case is, of course, flat Newton--Cartan spacetime \cite{Hartong:2014oma,Hartong:2015wxa}.
For suitable coordinates $x^0=t,\,x^1=x$ it is given by
\begin{equation}\label{eq:flat_gauge}
\tau_{\mu}  =  \delta_{\mu}^{t}, \qquad
e_{\mu}  =  \delta_{\mu}^{x}, \qquad
v^{\mu}  =  -\delta_{t}^{\mu}, \qquad
e^{\mu} = \delta_{x}^{\mu},\qquad
m_{\mu}  =  \partial_{\mu}\theta,
\end{equation}
where $\theta$ is an arbitrary function, i.e. $m_\mu$ is a pure gauge.
The residual coordinate transformations are exactly global Galilean transformations
\begin{align}
t^\prime &= t + a ,\\
x^\prime &= x+v t + b,\\
\theta^\prime (t^\prime,x^\prime)&= \theta(t,x) - \frac{1}{2} v^2 t + vx ,
\end{align}
where we have the parameters $v$ for the Galilean boost, and $a,b$ for the translations.
In particular we may choose $\theta=\mathrm{constant}$, but it will regain spacetime dependence after doing a boost.

\subsection{Covariant derivatives and matter actions}
If we want to geometrize a given non-relativistic field theory, we must use Galilean covariant derivatives for the theory to be compatible with local Galilei transformations.
Analogous to the relativistic case, we may introduce a covariant derivative $\nabla_\mu$ with a Galilean or Newton--Cartan affine connection $\Gamma_{\mu\nu}^\rho$ and a spin connection $\omega_\mu{}^A{}_B$.
The construction is similar to Lorentzian geometry, with the main difference being that the symmetry of the tangent and frame bundles is Galilean \cite{Bekaert:2014bwa,Bekaert:2015xua,Figueroa-OFarrill:2020gpr,Hansen:2020wqw}.
For this paper, it is sufficient to restrict to scalar fields $\phi$ where it, of course, reduces to the partial derivative, $\nabla_\mu\phi = \partial_\mu\phi$.
The general case is discussed further in Appendix \ref{sec:general_connections_curvatures}.

Besides the local Galilean symmetry, the scalar fields also transform under $\mathrm{U}(1)$ mass or particle number symmetry.
To not break the particle number symmetry, we must use a $\mathrm{U}(1)$ covariant derivative.
This is naturally defined by taking
\begin{equation}\label{eq:covariant_der_2d}
    D_\mu = \d_\mu + mm_\mu,
\end{equation}
where $m$ is the mass of the field.
A temporal derivative is then formed in a covariant way as $\sim v^\mu D_\mu$ and a spatial derivative as $\sim e^\mu D_\mu$.
Notice that the first transform under a local Galilean boost, which must be compensated by other terms.

Given a matter theory with field $\varphi$ coupled to a Newton--Cartan geometry, we can write its Bargmann invariant action $S$ and Lagrangian $\mathcal{L}$ as
\begin{equation}\label{eq:action_matter_NC}
    S\left[\varphi,\tau,e,m\right] = \int_M \der^2 x e\, \mathcal{L}\left[\varphi,\tau,e,m\right],
\end{equation}
where we have the Galilean invariant measure
\begin{equation}
    e:= \det(\tau_\mu,e_\mu).
\end{equation}
All dependence on $m_\mu$ must be through the covariant derivative.

In general, it is not an easy task to form Lagrangians that respect both boost and particle number symmetries \cite{Geracie:2014nka,Jensen:2014aia,Geracie:2016inm}.
One way where this is guaranteed is to consider \emph{null reductions} of relativistic theories: 
Any 3d field theory on a Lorentzian background with a null Killing vector can be null-reduced to a 2d field theory on the type of Newton--Cartan background we have considered here.
In particular, Schr\"odinger-type models are obtained by null reductions of Klein--Gordon-type actions.
In Appendix \ref{sec:null_reductions}, we review this procedure in more detail.

\subsection{Matter currents}
We can define three covariantly conserved currents of the action \eqref{eq:action_matter_NC} as the response to the variation of the background geometry:
\begin{equation}
\delta_{\mathrm{bgd}}S\left[\varphi,\tau,e,m\right]:=\int_{M}\mathrm{d}^{2}x\,e\left(\mathcal{E}^{\mu}\delta\tau_{\mu}+\mathcal{P}^{\mu}\delta e_{\mu}+\mathcal{J}^{\mu}\delta m_{\mu}\right)\label{eq:currents_typeI0_2d}
\end{equation}
or equivalently
\begin{eqnarray}\label{eq:currents_typeI1_2d}
\mathcal{E}^{\mu} := e^{-1}\frac{\delta S}{\delta\tau_{\mu}},\qquad
\mathcal{P}^{\mu} := e^{-1}\frac{\delta S}{\delta e_{\mu}},\qquad
\mathcal{J}^{\mu} := e^{-1}\frac{\delta S}{\delta m_{\mu}},
\end{eqnarray}
where here $\mathcal{E}^{\mu}$ is the energy current, $\mathcal{P}_{}^{\mu}$ the momentum current and $\mathcal{J}^{\mu}$ the mass current. These three currents are of fundamental importance to our constructions below.
We stress again that the mass current is \emph{always} present in a Bargmann invariant theory.
In fact, breaking the $\mathrm{U}(1)$ symmetry would lead to drastic consequences.
The particles' mass would no longer be well-defined, and the non-relativistic dispersion relation $E=P^2/(2m)$ would not hold. Such a situation is rather unphysical.
In quantum mechanics, we would furthermore have that the states are not localizable \cite{Bargmann:1954gh}.
In conclusion, we better not break the Bargmann symmetry to Galilei if we want to avoid odd physics.

The diffeomorphisms of the background fields \eqref{eq:trafo_NC_fields_1}-\eqref{eq:trafo_NC_fields_4} give the covariant conservation laws of the currents:
\begin{equation} \label{eq:consv_equation_2d}
0 =  \partial_{\mu}\mathcal{E}^{\mu}+\left(e^{-1}\partial_{\mu}e\right)\mathcal{E}^{\mu},\qquad
0 =  \partial_{\mu}\mathcal{P}^{\mu}+\left(e^{-1}\partial_{\mu}e\right)\mathcal{P}^{\mu},\qquad
0 = \partial_{\mu}\mathcal{J}^{\mu}+\left(e^{-1}\partial_{\mu}e\right)\mathcal{J}^{\mu}.
\end{equation}
These can be rewritten in terms of the covariant derivative $\nabla_\mu$ using $\left(e^{-1}\partial_{\mu}e\right)=\Gamma^\rho_{\mu\rho}$ for any Newton--Cartan connection $\Gamma^\rho_{\mu\nu}$.
We then obtain for $\mathcal{C}^\mu=\{\mathcal{E}^{\mu}, \mathcal{P}^{\mu}, \mathcal{J}^{\mu} \}$
\begin{equation}
0 = \nabla_{\mu}\mathcal{C}^{\mu}+2\Gamma^\rho_{[\mu\rho]}\mathcal{C}^{\mu},\label{eq:consv_equation_2d_cov}
\end{equation}
where $2\Gamma^\rho_{[\mu\rho]}$ is the torsion of the chosen connection.
Unlike the Levi-Civita connection, Newton--Cartan connections are naturally torsionful and non-metric compatible \cite{Hansen:2020wqw}.
More discussion can be found in Appendix \ref{sec:general_connections_curvatures}.

Finally, as the fields are not manifestly boost-invariant, a Galilean boost relates $\mathcal{P}^{\mu}$ and $\mathcal{J}^{\mu}$ through the on-shell Ward identity
\begin{equation}
\mathcal{P}^{\mu}\tau_{\mu}=-\mathcal{J}^{\mu}e_{\mu},
\end{equation}
which shows that $\mathcal{P}^{\mu}$ should be thought of as the stress-mass current.
This relation is well-known for non-relativistic theories on flat spacetimes.
Here it shows up as relations between the Noether currents after a simplifying redefinition \cite{Festuccia:2016awg}.

\section{Bilinear deformations and Newton--Cartan gravity}
\label{sec:bilinear}
This section defines three fundamental bilinear deformations for non-relativistic QFTs and derives the corresponding gravity formulations.
As reviewed in the previous section, any 2d non-relativistic Bargmann QFT has three fundamental symmetries, corresponding to mass, momentum, and energy conservation. The conserved currents are denoted by $\mathcal{J}^{\mu}$, $\mathcal{P}^{\mu}$ and $\mathcal{E}^{\mu}$, respectively, as reviewed in the previous section.
We can choose any two of them and construct a bilinear operator. For instance, the $\rT\overline{\rT}$ operator corresponds to the choice
\begin{align}
O_{2,1}=e\,\epsilon_{\mu\nu}\mathcal{E}^{\mu}\mathcal{P}^{\nu}=e\,\det(\mathcal{E}^{\mu},\mathcal{P}^{\nu}),
\end{align}
where $\epsilon_{\mu\nu}$ is the Levi-Civita symbol. The bilinear operator which triggers the hard rod deformation is $O_{0,1}=e\,\epsilon_{\mu\nu}\mathcal{J}^{\mu}\mathcal{P}^{\nu}$. The remaining bilinear operator is $O_{0,2}=e\,\epsilon_{\mu\nu}\mathcal{J}^{\mu}\mathcal{E}^{\nu}$. We shall call the corresponding deformation the $\rJ\rE$ deformation. The three bilinear deformations are defined by
\begin{align}
\label{eq:bilinear}
\frac{\rd S_{\lambda}}{\rd\lambda}=-\int \rd^2x\,e\, O_{a,b}(x),
\end{align}
where $S_{\lambda}$ is the deformed action and $(a,b)=(0,1),(0,2),(2,1)$. 

To derive the gravity formulation, we use the fact that the three conserved currents can be written as the variations of the action as defined through \eqref{eq:currents_typeI1_2d}.
Using these relations, the definition (\ref{eq:bilinear}) can be seen as equations for the classical action.
Let us explain this point more explicitly by the $\rT\overline{\rT}$ deformation. Consider the deformed partition function
\begin{align}
Z_{\lambda}=\int\mathcal{D}\phi\,e^{-S_{\lambda}[\phi]}.
\end{align}
From (\ref{eq:bilinear}), we have
\begin{align}
\label{eq:diffEq}
\frac{\rd Z_{\lambda}}{\rd\lambda}
=\int\mathcal{D}\phi \left(-\frac{\rd S_{\lambda}[\phi]}{\rd\lambda}\right)e^{-S{_\lambda}[\phi]}=\widehat{D}_{2,1}Z_{\lambda},
\end{align}
where
\begin{align}
\label{eq:op}
\widehat{D}_{2,1}=\int d^2x\,\epsilon_{\mu\nu}\,:\!\frac{\delta}{\delta\tau_{\mu}(x)}\frac{\delta}{\delta e_{\mu}(x)}\!:.
\end{align}
Here the normal ordering means we subtract the contribution of the term proportional to $\delta^2 S/\delta\tau_{\mu}\delta e_{\nu}$\footnote{More explicitly, we have $:\!\frac{\delta^2}{\delta\tau_{\mu}(x)\delta e_{\mu}(x)}e^{-S_{\lambda}}\!:=\frac{\delta^2}{\delta\tau_{\mu}(x)\delta e_{\mu}(x)}e^{-S_{\lambda}}+\frac{\delta^2 S_{\lambda}}{\delta\tau_{\mu}(x)\delta e_{\mu}(x)}e^{-S_{\lambda}}$.}. Similar equations can be derived for the other two deformations. Notice that (\ref{eq:diffEq}) is similar to a diffusion equation. We can solve it formally by a heat kernel-like approach, which will be discussed shortly. The formal solution gives us the gravity action immediately. This approach has been applied in the relativistic QFTs in \cite{Cottrell:2018skz,Mazenc:2019cfg,Aguilera-Damia:2019tpe}.

\subsection{Simple examples}
As a warm-up, we consider two simple examples. The first one is the 1d heat equation
\begin{align}
\partial_tf(t,x)=\partial_x^2f(t,x),
\end{align}
which can be solved in a few steps: First, we can write
\begin{align}
\label{eq:formalF}
f(t,x)=e^{t\partial_x^2}f(0,x).
\end{align}
Secondly, rewrite
\begin{align}
\label{eq:secondSte}
f(0,x)=\int\rd y\,\delta(x-y)f(0,y)=\frac{1}{2\pi}\int\rd y\int\rd p\, e^{ip(x-y)}f(0,y).
\end{align}
Finally, plug (\ref{eq:secondSte}) into the right-hand side of (\ref{eq:formalF}) and integrate out $p$. We obtain
\begin{align}
f(t,x)=\frac{1}{2\sqrt{\pi t}}\int\rd y\, e^{-\frac{1}{4t}(x-y)^2}f(0,y).
\end{align}
This is nothing but the heat kernel solution.\par

Next, we consider a slightly more non-trivial equation which involves functional derivatives\footnote{The interpretation of normal ordering is the same as (\ref{eq:op}).}
\begin{align}
\partial_t Z_{t}[\phi]=\int\rd^2 x\,:\!\frac{\delta}{\delta\phi(x)}\frac{\delta}{\delta\phi(x)}\!:\,Z_{t}[\phi].
\end{align}
This equation can be solved similarly. First, we have
\begin{align}
\label{eq:formalZ}
Z_t[\phi]=\exp\left[t\int\rd^2x\,:\!\frac{\delta}{\delta\phi(x)}\frac{\delta}{\delta\phi(x)}\!: \right]Z_0[\phi].
\end{align}
Secondly, rewrite
\begin{align}
Z_0[\phi]=\int\mathcal{D}\varphi\,\delta(\phi-\varphi)Z_0[\varphi]
\propto\int\mathcal{D}\varphi\int\mathcal{D}J\,e^{\int\rd^2 xJ(x)[\phi(x)-\varphi(x)]}Z_0[\varphi],
\end{align}
where $\delta(\phi-\varphi)$ is the functional delta-function in the proper sense and $\varphi$ is an auxiliary scalar field to be integrated over. Plugging into (\ref{eq:formalZ}) and integrating out $J(x)$, we obtain
\begin{align}
Z_t[\phi]\propto\int\mathcal{D}\varphi\,e^{-\frac{1}{4t}\int\rd^2x[\phi(x)-\varphi(x)]^2}Z_0[\varphi],
\end{align}
where we have neglected the prefactors which are not important for us.

\subsection{The bilinear deformations}
We can apply the same method to solve the flow equation of the bilinear deformations (\ref{eq:bilinear}). Let us consider the $\rT\overline{\rT}$ deformed partition function
\begin{align}
\partial_{\lambda}Z_{\lambda}[\tau_{\mu},e_{\mu}]=\widehat{D}_{2,1}Z_{\lambda}[\tau_{\mu},e_{\mu}].
\end{align}
Following the same steps, we arrive at the following formal solution
\begin{align}
\label{eq:solZlambda}
Z_{\lambda}[\tau_{\mu},e_{\mu}]
=\int\mathcal{D}\tilde{\tau}_{\mu}\mathcal{D}\tilde{e}_{\mu}\exp\left[-\frac{1}{\lambda}\int\rd^2 x\epsilon^{\mu\nu}(\tau_{\mu}-\tilde{\tau}_{\mu})(e_{\nu}-\tilde{e}_{\nu}) \right]Z_0[\tilde{\tau}_{\mu},\tilde{e}_{\mu}],
\end{align}
where as before we neglected the prefactors from the Gaussian integral.
From (\ref{eq:solZlambda}), we can extract the deformed classical action
\begin{align}
\label{eq:TTbar}
S_{\lambda}^{\rT\overline{\rT}}[\tau_{\mu},e_{\mu};\tilde{\tau}_{\mu},\tilde{e}_{\mu}|\phi]=
\frac{1}{\lambda}\int\rd^2 x\,\epsilon^{\mu\nu}(\tau_{\mu}-\tilde{\tau}_{\mu})(e_{\nu}-\tilde{e}_{\nu})+S_0[\tilde{\tau}_{\mu},\tilde{e}_{\mu}|\phi].
\end{align}
Here $S_0$ is the undeformed action on the background described by the auxiliary zweibein $(\tilde{\tau}_{\mu},\tilde{e}_{\mu})$. The other term can be interpreted as the non-relativistic gravity action, which couples to the undeformed theory. Notice that the gravity action takes the same form as its relativistic counterpart \cite{Dubovsky:2018bmo}, but it is also manifestly Galilean invariant under \eqref{eq:trafo_NC_fields_1}-\eqref{eq:trafo_NC_fields_4}.\par

Similarly, for the hard rod and JE deformations, we find the deformed classical actions
\begin{align}
\label{eq:SHR}
S_{\lambda}^{\text{HR}}[m_{\mu},e_{\mu};\tilde{m}_{\mu},\tilde{e}_{\mu}|\phi]=
\frac{1}{\lambda}\int\rd^2 x\,\epsilon^{\mu\nu}(m_{\mu}-\tilde{m}_{\mu})(e_{\nu}-\tilde{e}_{\nu})+S_0[\tilde{m}_{\mu},\tilde{e}_{\mu}|\phi]
\end{align}
and
\begin{align}
\label{eq:SJE}
S_{\lambda}^{\text{JE}}[m_{\mu},\tau_{\mu};\tilde{m}_{\mu},\tilde{\tau}_{\mu}|\phi]=
\frac{1}{\lambda}\int\rd^2 x\,\epsilon^{\mu\nu}(m_{\mu}-\tilde{m}_{\mu})(\tau_{\nu}-\tilde{\tau}_{\nu})+S_0[\tilde{m}_{\mu},\tilde{\tau}_{\mu}|\phi].
\end{align}
Notice that for these two deformations, the gravity action involve both the zweibein and the $\mathrm{U}(1)$ gauge field, similar to those of $JT_a$ deformed QFTs \cite{Aguilera-Damia:2019tpe,Anous:2019osb}.

Since our derivation of the deformed action is somewhat heuristic, let us now verify that the deformed actions indeed satisfy the definition of the bilinear deformations (\ref{eq:bilinear}). We consider the $\rT\overline{\rT}$ deformation as an example. The proof for the other two cases is similar.\par

The action (\ref{eq:TTbar}) depends on both the zweibein $(\tau_{\mu},e_{\mu})$ and the auxiliary zweibein $(\tilde{\tau}_{\mu},\tilde{e}_{\mu})$. To obtain the $\rT\overline{\rT}$ deformed classical action, we need to integrate out the auxiliary zweibein $(\tilde{\tau}_{\mu},\tilde{e}_{\mu})$. In practice, this means finding the saddle-point solution for $(\tilde{\tau}_{\mu},\tilde{e}_{\mu})$ and plug it back in. We denote the saddle-point solution by $\tilde{\tau}_{\mu}^{\star}$ and $\tilde{e}_{\mu}^{\star}$. We will show that $S_{\rT\overline{\rT}}[\tau_{\mu},e_{\mu}|\phi]:= S_{\lambda}^{\rT\overline{\rT}}[\tau_{\mu},e_{\mu};\tilde{\tau}_{\mu}^{\star},\tilde{e}_{\mu}^{\star}|\phi]$ satisfies the definition of the $\rT\overline{\rT}$ deformation. Taking variation of (\ref{eq:TTbar}) with respect to $\tilde{\tau}_{\mu}$ and $\tilde{e}_{\mu}$, we obtain the saddle-point equations
\begin{align}
\tau_{\mu}=\tilde{\tau}_{\mu}^{\star}+\lambda\,\tilde{e}\,\epsilon_{\mu\nu}\,\mathcal{P}_0^{\nu},\qquad e_{\mu}=\tilde{e}_{\mu}^{\star}-\lambda\,\tilde{e}\,\epsilon_{\mu\nu}\mathcal{E}^{\nu}_0,
\end{align}
where $\mathcal{P}_0$ and $\mathcal{E}_0$ are the undeformed currents. On the other hand, taking variations with respect to $\tau_{\mu}$ and $e_{\mu}$ lead to the definitions of the deformed currents
\begin{align}
\frac{\delta S_{\rT\overline{\rT}}[\tau_{\mu},e_{\mu}|\phi]}{\delta \tau_{\mu}}=e\,\mathcal{E}^{\mu},\qquad
\frac{\delta S_{\rT\overline{\rT}}[\tau_{\mu},e_{\mu}|\phi]}{\delta e_{\mu}}=e\,\mathcal{P}^{\mu},
\end{align}
which can be written as
\begin{align}
\label{eq:defCurrent}
\tau_{\mu}=\tilde{\tau}_{\mu}+\lambda\, {e}\,\epsilon_{\mu\nu}\,\mathcal{P}^{\nu},\qquad e_{\mu}=\tilde{e}_{\mu}-\lambda\, {e}\,\epsilon_{\mu\nu}\mathcal{E}^{\nu}.
\end{align}
Here $\mathcal{E}^{\mu}$ and $\mathcal{P}^{\mu}$ are the deformed currents. Taking derivative of $S_{\rT\overline{\rT}}$ with respect to $\lambda$, we have
\begin{align}
\label{eq:derive}
\frac{\rd S_{\rT\overline{\rT}}}{\rd\lambda}=-\frac{1}{\lambda^2}\int\rd^2x\,\epsilon^{\mu\nu}
(\tau_{\mu}-\tilde{\tau}_{\mu}^{\star})(e_{\nu}-\tilde{e}_{\nu}^{\star})
+\frac{\delta S_{\rT\overline{\rT}}}{\delta\tilde{\tau}_{\mu}^{\star}}\frac{\partial\tilde{\tau}_{\mu}^{\star}}{\partial\lambda}
+\frac{\delta S_{\rT\overline{\rT}}}{\delta\tilde{e}_{\mu}^{\star}}\frac{\partial\tilde{e}_{\mu}^{\star}}{\partial\lambda}.
\end{align}
From the definition of saddle-point equation, 
\begin{align}
\frac{\delta S_{\rT\overline{\rT}}}{\delta\tilde{\tau}_{\mu}^{\star}}=\frac{\delta S_{\rT\overline{\rT}}}{\delta\tilde{e}_{\mu}^{\star}}=0,
\end{align}
and therefore the last two terms in (\ref{eq:derive}) vanish. Using (\ref{eq:defCurrent}), we have
\begin{align}
\frac{\rd S_{\rT\overline{\rT}}}{\rd\lambda}=-\int\rd^2x\,e^2\,\epsilon_{\mu\nu}\mathcal{E}^{\mu}\mathcal{P}^{\nu}=-\int\rd^2x\,e\,O_{2,1},
\end{align}
which is precisely (\ref{eq:bilinear}).
The generalization to the other two cases is straightforward.

\section{Deformed classical Lagrangians}
\label{sec:defLag}
In this section, we derive the deformed classical Lagrangian for the Schr\"odinger model with a generic potential for all three bilinear deformations. The same result can be derived from two different approaches. The first one is the direct approach where one starts from the definition (\ref{eq:bilinear}) and work out the deformed Lagrangian order by order in $\lambda$; the other approach exploits the gravity interpretation we developed in the previous section and gives closed-form expressions for the deformed Lagrangians. The fact that these two quite different methods lead to the same deformed Lagrangians can be seen as a non-trivial test of our proposal.

\subsection{Schr\"odinger model and conserved currents}
Let us first define the Schr\"odinger model and its conserved currents. In curved space, the action is given by
\begin{equation}\label{eq:schr_2d_action}
    S_{\text{Sch}}=\int{\rm d}^{2}x\,e\,\left[\frac{i}{2}v^{\mu}(\phi D_{\mu}\phi^{\dagger}-\phi^{\dagger}D_{\mu}\phi)-e^{\mu}e^{\nu}
    D_{\mu}\phi^{\dagger}D_{\nu}\phi-V(|\phi|) \right],
\end{equation}
where $\phi$ is a complex scalar field and $V(|\phi|)$ is any potential that does not depend on the metric nor on the covariant derivative \eqref{eq:covariant_der_2d} with $m=1/2$. Furthermore, we require that the potential is invariant under Hermite conjugation. Taking $V(|\phi|)=0$ leads to the non-relativistic free boson. One slightly more non-trivial example is the Lieb--Liniger model, or the non-linear Schr\"odinger model where we take $V(|\phi|)=c\,\phi^{\dagger}\phi^{\dagger}\phi\phi$, with $c$ being the coupling constant. One nice feature of the Schr\"odinger model is that it can be obtained from a relativistic 3d Klein--Gordon type theory. This is described in detail in
Appendix~\ref{sec:null_reductions}.

In flat spacetime, we recover the familiar Schr\"odinger action
\begin{align}
\label{eq:flatS}
S=\int\rd^2x\,\mathcal{L}=\int\rd^2x\left[\frac{i}{2}(\phi^{\dagger}\partial_t\phi-\phi\partial_t\phi^{\dagger})-\partial_x\phi^{\dagger}\partial_x\phi
-V(|\phi|) \right].
\end{align}
Now we discuss the symmetries of the action \eqref{eq:flatS}.
Spacetime translation invariance leads to a conserved local Noether stress-energy tensor $T^{\mu}{}_{\nu}$. In terms of the Lagrangian,
\begin{align}
T^{\mu}{}_{\nu}=\frac{\partial\mathcal{L}}{\partial(\partial_{\mu}\phi)}\partial_{\nu}\phi
+\frac{\partial\mathcal{L}}{\partial(\partial_{\mu}\phi^{\dagger})}\partial_{\nu}\phi^{\dagger}-\mathcal{L}\,\delta_{\nu}^{\mu}.
\end{align}
The energy and momentum currents are identified with $\mathcal{E}^{\mu}=T^{\mu}{}_{t}$ and $\mathcal{P}^{\mu}=T^{\mu}{}_{x}$, respectively.
In addition, there is a global $\mathrm{U}(1)$ symmetry $\phi\mapsto e^{i\theta}\phi$ which is related to the conservation of mass. The Noether current is given by
\begin{align}
\label{eq1}
\mathcal{J}^{\mu} =\frac{i}{2}\left(\frac{\partial\mathcal{L}}{\partial(\partial_{\mu}\phi)}\phi
-\frac{\partial\mathcal{L}}{\partial(\partial_{\mu}\phi^{\dagger})}\phi^{\dagger}\right).
\end{align}
The corresponding conserved charge is the total mass of the system. The three conserved currents written explicitly are
\begin{align}
\label{eq:flatCurrent}
\mathcal{E}^{\mu}=&\,\frac{i}{2}\delta^{\mu}_t\left(\phi\partial_t\phi^{\dagger}-\phi^{\dagger}\partial_t\phi-2i\mathcal{L}\right)
+\delta_x^{\mu}\,\left(\partial_x\phi^{\dagger}\partial_t\phi+\partial_x\phi\partial_t\phi^\dagger\right),\\\nonumber
\mathcal{P}^{\mu}=&\,\frac{i}{2}\delta^{\mu}_t\left(\phi\partial_x\phi^{\dagger}-\phi^{\dagger}\partial_x\phi\right)
+\delta^{\mu}_x\left(\mathcal{L}+2\partial_x\phi\partial_x\phi^{\dagger} \right),\\\nonumber
\mathcal{J}^{\mu}=&\,-\frac{1}{2}\delta^{\mu}_t\,\phi\phi^{\dagger}
+\frac{i}{2}\delta^{\mu}_x\left(\phi^{\dagger}\partial_x\phi-\phi\partial_x\phi^{\dagger} \right).
\end{align}
In general curved background, the currents can be calculated via \eqref{eq:currents_typeI1_2d}, leading to
\begin{align}
\label{eq:curvedCurrent}
\mathcal{E}^{\mu}=&\,\frac{i}{2}v^{\mu}v^{\rho}\left(\phi D_{\rho}\phi^{\dagger}-\phi^{\dagger}D_{\rho}\phi\right)
-e^{\mu}e^{\rho}v^{\sigma}\left(D_{\rho}\phi^{\dagger}D_{\sigma}\phi+D_{\sigma}\phi^{\dagger}D_{\rho}\phi \right)-v^{\mu}\,\mathcal{L},\\\nonumber
\mathcal{P}^{\mu}=&\,\frac{i}{2}v^{\mu}e^{\rho}\left(\phi^{\dagger}D_{\rho}\phi-\phi D_{\rho}\phi^{\dagger}\right)
+e^{\mu}e^{\rho}e^{\sigma}\left(D_{\rho}\phi^{\dagger}D_{\sigma}\phi+D_{\sigma}\phi^{\dagger}D_{\rho}\phi \right)+e^{\mu}\,\mathcal{L},\\\nonumber
\mathcal{J}^{\mu}=&\,\frac{1}{2}v^{\mu}\,\phi\phi^{\dagger}+\frac{i}{2}e^{\mu}e^{\nu}\left(\phi^{\dagger}D_{\nu}\phi-\phi D_{\nu}\phi^{\dagger} \right).
\end{align}
It is easy to check that on flat spacetime as described in section \ref{sec:NC_flat_spacetime}, \eqref{eq:curvedCurrent} reduces to \eqref{eq:flatCurrent}.

\subsection{Deformed Lagrangian I. The direct approach}
We are now ready to derive the deformed Lagrangians for the three bilinear deformations. In this subsection, we perform the calculation using the direct approach. This was the first approach to derive the deformed Lagrangian in the relativistic case (see for example \cite{Cavaglia:2016oda,Bonelli:2018kik}). In this approach, one performs a formal expansion of the deformed Lagrangian
\begin{align}
\mathcal{L}_{\lambda}=\sum_{k=0}^{\infty} L_k\,\lambda^k
\end{align}
and calculates $L_k$ order by order using the definition. This can always be worked out explicitly up to certain orders in $\lambda$. By observing the patterns of $L_k$, one can usually make an ansatz for the deformed Lagrangian and turn the definition into a differential equation, which can be solved and gives the deformed Lagrangian. This strategy works fairly well for the hard rod deformation because the deformed Lagrangian takes a compact form, and it is relatively easy to guess the general pattern. However, for $\rT\overline{\rT}$ and JE deformations, the closed-form deformed Lagrangians are quite complicated, as we will see shortly. Therefore it is hard to see the patterns in these cases. Nevertheless, we can work out the results up to relatively high orders as perturbative data, which can be checked against the closed-form expressions obtained from other approaches.
\paragraph{The hard rod deformation} Let us start with the simplest case, namely the hard rod deformation. We work in a flat spacetime. The definition is
\begin{equation}
\label{eq:Ldef01}
\frac{{\rm d}\mathcal{L}_{\lambda}^{\text{HR}}}{{\rm d}\lambda}=-\epsilon_{\mu\nu}\mathcal{J}^{\mu}\mathcal{T}^{\nu}.
\end{equation}
We expand both the Lagrangian and current densities in $\lambda$
\begin{align}
\mathcal{L}_{\lambda} =\sum_{n=0}^{\infty}L_n \lambda^{n},\qquad 
\mathcal{J}^{\mu} =\sum_{n=0}^{\infty}J_{n}^{\mu}\lambda^{n},\qquad\mathcal{P^{\mu}}=\sum_{n=0}^{\infty}P_{n}^{\mu}\lambda^{n}.
\end{align}
Plugging into (\ref{eq:Ldef01}), we obtain the following recursion relation
\begin{equation}
\label{eq:recurs01}
L_{n+1}=-\frac{1}{n+1}\sum_{k=0}^{n}\epsilon_{\mu\nu}J_{k}^{\mu}P_{n-k}^{\nu},
\end{equation}
where $J_{k}^{\mu}$ and $T_{k}^{\mu}$ are defined via $L_k$:
\begin{align} 
\label{eq3}
J_{k}^{\mu}:=&\,i\left(\frac{\partial L_k}{\partial(\partial_{\mu}\phi)}\phi-
\frac{\partial L_k}{\partial(\partial_{\mu}\phi^{\dagger})}\phi^{\dagger}\right),\\\nonumber
P_{k}^{\mu}:=&\,\frac{\partial L_k}{\partial(\partial_{\mu}\phi)}\partial_{x}\phi+
\frac{\partial L_k}{\partial(\partial_{\mu}\phi^{\dagger})}\partial_{x}\phi^{\dagger}-L_k\,\delta_{x}^{\mu}.
\end{align}
Using (\ref{eq3}), (\ref{eq:recurs01}) and the initial condition $L_0=\mathcal{L}$ in (\ref{eq:flatS}), we can calculate $L_k$ order by order. The first few orders are given by
\begin{align}
L_1=&\,\left(-\frac{\phi\phi^{\dagger}}{2}\right)\left(\mathcal{L}+\phi_x\phi_x^{\dagger}\right)
-\frac{1}{4}\left(\phi^2(\phi_x^{\dagger})^2+\phi_x^2(\phi^{\dagger})^2\right),\\\nonumber
L_2=&\,\left(-\frac{\phi\phi^{\dagger}}{2}\right)^2\left(\mathcal{L}+\phi_x\phi_x^{\dagger}\right)+\frac{1}{16}\phi\phi^{\dagger}
(\phi_x^{\dagger}\phi-\phi^{\dagger}\phi_x)^2,\\\nonumber
L_3=&\,\left(-\frac{\phi\phi^{\dagger}}{2}\right)^3\left(\mathcal{L}+\phi_x\phi_x^{\dagger}\right)-\frac{1}{32}(\phi\phi^{\dagger})^2
(\phi_x^{\dagger}\phi-\phi^{\dagger}\phi_x)^2.
\end{align}
where we have defined the shorthand notations
\begin{align}
\phi_t:=\partial_t\phi,\qquad \phi_x:=\partial_x\phi,\qquad \phi_t^{\dagger}:=\partial_t\phi^{\dagger},\qquad\phi_x^{\dagger}:=\partial_x\phi^{\dagger}.
\end{align}
Working out a few more orders, we can find the pattern
\begin{align}
L_n=\left(-\frac{\phi\phi^{\dagger}}{2}\right)^n(\mathcal{L}+\phi_x\phi^{\dagger}_x)+\frac{(-1)^n}{2^{n+2}}(\phi\phi^{\dagger})^{n-1}(\phi_x^{\dagger}\phi-\phi^{\dagger}\phi_x),\qquad n\ge 2.
\end{align}
The full deformed Lagrangian is then given by
\begin{align}
\label{eq:closedFHR}
\mathcal{L}_{\lambda}^{\text{HR}}=\frac{1}{2+\lambda\phi\phi^{\dagger}}\left(2\mathcal{L}-\frac{\lambda}{8}(4+\lambda\phi\phi^{\dagger})
(\phi_x^{\dagger}\phi+\phi_x\phi^{\dagger})^2\right).
\end{align}

\paragraph{The $\rT\overline{\rT}$ deformation} Now we consider the $\rT\overline{\rT}$ deformation whose definition is
\begin{equation}
\frac{{\rm d}\mathcal{L}_{\lambda}}{{\rm d}\lambda}=-\epsilon_{\mu\nu}\mathcal{E}^{\mu}\mathcal{P}^{\nu}.
\end{equation}
Similarly, we expand the Lagrangian and the currents in $\lambda$
\begin{align}
\mathcal{L}_{\lambda}=\sum_{n=0}^{\infty}L_n\,\lambda^{n},\qquad \mathcal{\mathcal{\mathcal{E}}^{\mu}}=\sum_{n=0}^{\infty}E_{n}^{\mu}\lambda^{n},\qquad \mathcal{P}^{\mu}=\sum_{n=0}^{\infty}P_{n}^{\mu}\lambda^{n},
\end{align}
which leads to a similar recursion relation
\begin{equation}
L_{n+1}=-\frac{1}{n+1}\sum_{k=0}^{n}\epsilon_{\mu\nu}E_{k}^{\mu}P_{n-k}^{\nu},
\end{equation}
where
\begin{align}
\label{eq:pertE}
E_{k}^{\mu}:=\frac{\partial L_{k}}{\partial(\partial_{\mu}\phi)}\partial_{t}\phi+
\frac{\partial{L}_{k}}{\partial(\partial_{\mu}\phi^{\dagger})}\partial_{t}\phi^{\dagger}-{L}_{k}\delta_{t}^{\mu}
\end{align}
and $P_k^{\mu}$ is given in (\ref{eq3}). The first few orders are $L_0=\mathcal{L}$ and
\begin{align}
\label{eq:pertLTTbar}
L_1=&\,\frac{1}{2}\left(i\phi\phi_{t}\phi_{x}^{\dagger2}-i\phi^{\dagger}\phi_{t}^{\dagger}\phi_{x}^{2}\right)+\phi_{x}^{\dagger2}\phi_{x}^{2}
+\frac{i}{2}(\phi^{\dagger}\phi_{t}-\phi_{t}^{\dagger}\phi)\mathcal{V}-\mathcal{V}^{2},\\\nonumber
L_2=&\,-\mathcal{V}^{3}+\frac{i}{2}(\phi^{\dagger}\phi_{t}-\phi\phi_{t}^{\dagger})\mathcal{V}^{2}
-\phi_{x}^{\dagger2}\phi_{x}^{2}\,\mathcal{V}-2\phi_x^3\phi_x^{\dagger3}+i\phi_x\phi_x^{\dagger}\left(\phi_t^{\dagger}\phi_x^2\phi^{\dagger}-\phi_t\phi_x^{\dagger2}\phi \right)\\\nonumber
&\,-\frac{1}{2}\phi_t\phi_t^{\dagger}\phi_x\phi_x^{\dagger}\phi\phi^{\dagger}+\frac{i}{2}\phi_x^2\phi_x^{\dagger2}\left(\phi_t\phi^{\dagger}
- \phi_t^{\dagger}\phi\right)+\frac{1}{4}\phi_t\phi_t^{\dagger}\left(\phi_x^{\dagger2}\phi^2+\phi_x^2\phi^{\dagger2} \right).
\end{align}
We see that the second-order result is already quite lengthy. Higher-order terms are more complicated, and it is hard to see the general pattern. Nevertheless, we can derive a closed-form result from the gravity approach.

\paragraph{The JE deformation} This deformation is defined using the mass and energy currents
\begin{equation}
    \frac{{\rm d}\mathcal{L}^{\text{JE}}_{\lambda}}{{\rm d}\lambda}=-\epsilon_{\mu\nu}\mathcal{J}^{\mu}\mathcal{E}^{\nu}.
\end{equation}
We expand the Lagrangian and currents in $\lambda$ as before, which leads to the following recursion relation
\begin{equation}
L^{\text{JE}}_{n+1}=-\frac{1}{n+1}\sum_{k=0}^{n}\epsilon_{\mu\nu}J_{k}^{\mu}E_{n-k}^{\nu},
\end{equation}
where $J_{k}^{\mu}$ and $E_{k}^{\nu}$ are defined via $L_{k}$ in (\ref{eq3}) and (\ref{eq:pertE}). The first few $L_k$ are given by $L_0=\mathcal{L}$ and
\begin{align}
L_1=&\,\frac{i}{2}(\phi_x\phi^{\dagger}-\phi_x^{\dagger}\phi)\mathcal{V}+\frac{i}{2}\phi_x\phi^{\dagger}_x(\phi_x\phi^{\dagger}-\phi_x^{\dagger}\phi)
-\frac{1}{2}\phi\phi^{\dagger}(\phi_t\phi_x^{\dagger}+\phi_t^{\dagger}\phi_x),\\\nonumber
L_2=&\,-\frac{1}{4}\phi\phi^{\dagger}\mathcal{V}^2+\frac{1}{4}\left(\phi^2\phi_x^{\dagger2}+\phi^{\dagger2}\phi_x^2
-4\phi\phi^{\dagger}\phi_x\phi_x^{\dagger}+i\phi\phi^{\dagger}(\phi_t\phi^{\dagger}-\phi_t^{\dagger}\phi) \right)\mathcal{V}\\\nonumber
&\,+\frac{1}{4}\phi_x\phi_x^{\dagger}\left(\phi^2\phi_x^{\dagger2}+\phi^{\dagger2}\phi_x^2\right)
-\frac{1}{4}\phi_t\phi_t^{\dagger}\phi^2\phi^{\dagger2}-\frac{3}{4}\phi_x^2\phi_x^{\dagger2}\phi\phi^{\dagger}\\\nonumber
&\,+\frac{i}{2}\phi_x\phi_x^{\dagger}\phi\phi^{\dagger}\left(\phi_t\phi^{\dagger}-\phi_t^{\dagger}\phi\right)
-\frac{i}{4}\phi\phi^{\dagger}\left(\phi_t\phi_x^{\dagger2}\phi-\phi_t^{\dagger}\phi_x^2\phi^{\dagger}\right).
\end{align}
Like the $\rT\overline{\rT}$ deformation, the results get more involved at higher orders, and the pattern for a closed-form expression for $L_n$ is not obvious.

\subsection{Deformed Lagrangian II. The gravity approach}
In this subsection, we present another approach to compute the deformed classical Lagrangian. This approach exploits the gravity actions (\ref{eq:TTbar}), (\ref{eq:SHR}) and (\ref{eq:SJE}). We take the $\rT\overline{\rT}$ deformation as an example. Starting from (\ref{eq:TTbar}), we fix the zweibein $\tau_{\mu},e_{\mu}$ in the flat gauge $\tau_{\mu}=\delta_{\mu}^t$ and $e_{\mu}=\delta_{\mu}^x$ described in section \ref{eq:flat_gauge}. Then we find the saddle points $\tilde{\tau}_{\mu}^{\star}$ and $\tilde{e}_{\mu}^{\star}$ and plug back in the action. The deformed action is given by $S_{\rT\overline{\rT}}[\delta_{\mu}^t,\delta_{\mu}^x;\tilde{\tau}_{\mu}^{\star},\tilde{e}_{\mu}^{\star}|\phi]$. This approach has been applied to the relativistic case in \cite{Caputa:2020lpa}. The deformed actions for the other two deformations can be obtained similarly. The main difference is that the flat limit of the $\mathrm{U}(1)$ gauge field is a pure gauge $m_{\mu}=\partial_{\mu}\theta$. In this subsection, we can simply take $m_{\mu}=0$ by gauge fixing.

\paragraph{The hard rod deformation} We consider hard rod deformation first. The saddle-point equation obtained from (\ref{eq:SHR}) is
\begin{align}
\label{eq:saddleHR}
0=\tilde{m}_{\mu}+\lambda\,\tilde{e}\,\epsilon_{\mu\nu}\,\mathcal{P}^{\nu}_0,\qquad
\delta_{\mu}^x=\tilde{e}_{\mu}-\lambda\,\tilde{e}\,\epsilon_{\mu\nu}\,\mathcal{J}^{\nu}_0.
\end{align}
where $\mathcal{P}_0^{\nu}$ and $\mathcal{J}_0^{\nu}$ are given in (\ref{eq:curvedCurrent}). There is a unique solution to this equation, which is given by
\begin{align}
\tilde{m}_t^{\star}=&\,\frac{-\lambda}{2+\lambda\phi\phi^{\dagger}}\left(\frac{1}{8}\lambda(\phi\phi_x^{\dagger}+\phi^{\dagger}\phi_x)^2
(4+\lambda\phi^{\dagger}\phi)+2(\mathcal{L}_0+2\phi_x\phi_x^{\dagger})\right),\\\nonumber
\tilde{m}_x^{\star}=&\,\frac{i\lambda}{2+\lambda\phi\phi^{\dagger}}(\phi_x^{\dagger}\phi-\phi_x\phi^{\dagger}),\\\nonumber
\tilde{e}_t^{\star}=&\,\frac{i\lambda}{2+\lambda\phi\phi^{\dagger}}(\phi_x^{\dagger}\phi-\phi_x\phi^{\dagger}),\\\nonumber
\tilde{e}_x^{\star}=&\,\frac{2}{2+\lambda\phi\phi^{\dagger}}.
\end{align}
Plugging these into $S_{\lambda}^{\text{HR}}[0,\delta_{\mu}^x;\tilde{m}_{\mu}^{\star},\tilde{e}_{\mu}^{\star}|\phi]$ reproduces precisely the deformed Lagrangian given in (\ref{eq:closedFHR}).
Notice that the deformation gives a non-trivial Newtonian potential $\Phi^{\star} = \tilde{m}_t^{\star}$.

\paragraph{The $\rT\overline{\rT}$ deformation} Now we consider the $\rT\overline{\rT}$ deformation. The saddle-point equation reads
\begin{equation}
\delta_{\mu}^t=\tilde{\tau}_{\mu}+\lambda\,\tilde{e}\,\epsilon_{\mu\nu}\,\mathcal{P}^{\nu}_0,\qquad
\delta_{\mu}^x=\tilde{e}_{\mu}-\lambda\,\tilde{e}\,\epsilon_{\mu\nu}\,\mathcal{E}^{\nu}_0.
\end{equation}
These equations can also be solved explicitly. There are several solutions to this equation; we choose the one that is regular in the $\lambda\to0$ limit. However, the solution is too involved to be presented here explicitly. It can be found in the ancillary notebook. Plugging in $S_{\lambda}^{\rT\overline{\rT}}[\delta_{\mu}^t,\delta_{\mu}^x;\tilde{\tau}_{\mu}^{\star},\tilde{e}_{\mu}^{\star}|\phi]$, we obtain the following expression for the deformed Lagrangian
\begin{align}
\label{eq:closedTTbar}
\mathcal{L}_{\lambda}^{\rT\overline{\rT}}=-\frac{F_1+2\sqrt{F_2}}{4\lambda(1-\lambda\mathcal{V})},
\end{align}
where
\begin{align}
F_1=&\,i\lambda(\phi_t^{\dagger}\phi-\phi_t\phi^{\dagger})+4\lambda\,\mathcal{V}-2,\\\nonumber
F_2=&\,\lambda^4\phi\phi^{\dagger}(\phi_t^{\dagger}\phi_x-\phi_t\phi_x^{\dagger})\,\mathcal{V}
-\lambda^3(\phi_t^{\dagger}\phi_x-\phi_t\phi_x^{\dagger})\left(2i\mathcal{V}(\phi_x^{\dagger}\phi+\phi_x\phi^{\dagger})
+\phi\phi^{\dagger}(\phi_t^{\dagger}\phi_x-\phi_t\phi_x^{\dagger})\right)\\\nonumber
&\,+\lambda^2\left(\frac{1}{2}\phi_t\phi_t^{\dagger}\phi\phi^{\dagger}-4\phi_x\phi_x^{\dagger}\mathcal{V}
-\frac{1}{4}(\phi_t^{\dagger2}\phi^2+\phi_t^2\phi^{\dagger2})+2i(\phi_t^{\dagger}\phi_x-\phi_t\phi_x^{\dagger})
(\phi_x^{\dagger}\phi+\phi_x\phi^{\dagger}) \right)\\\nonumber
&\,+\lambda\left(4\phi_x^{\dagger}\phi_x+i(\phi_t^{\dagger}\phi-\phi_t\phi^{\dagger})\right)+1.
\end{align}
As we can see, the result is highly non-trivial. Performing a perturbative expansion in $\lambda$ to high powers, we can check that the results match what we obtained from the direct approach (\ref{eq:pertLTTbar}).

\paragraph{The JE deformation} Finally, we consider the JE deformation. The saddle-point equations read
\begin{align}
0=\tilde{m}_{\mu}+\lambda\,\tilde{e}\,\epsilon_{\mu\nu}\,\mathcal{E}_0^{\nu},\qquad
\delta_{\mu}^t=\tilde{\tau}_{\mu}-\lambda\,\tilde{e}\,\epsilon_{\mu\nu}\,\mathcal{J}_{0}^{\nu}.
\end{align}
The solution can be found but is again rather involved to be written down. There is a unique solution which is regular in the $\lambda\to 0$ limit. Plugging this solution to $S_{\lambda}^{\text{JE}}[0,\delta_{\mu}^t;\tilde{m}_{\mu}^{\star},\tilde{\tau}_{\mu}^{\star}|\phi]$, we obtain the deformed Lagrangian
\begin{equation}
\label{eq:closedformJE}
\mathcal{L}_{\lambda}^{\text{JE}}=\frac{G_{1}+\sqrt{G_{2}}}{2\lambda^{2}\phi^{\dagger}\phi},
\end{equation}
where
\begin{align}
G_{1}=&\,2i\lambda(\phi^{\dagger}_x\phi-\phi_x\phi^{\dagger})-4,\\
G_{2}=&\,-\lambda^4\phi^2\phi^{\dagger2}\left(\phi_t^{\dagger}\phi+\phi_t\phi^{\dagger}\right)
-4\lambda^3\phi\phi^{\dagger}(\phi_t^{\dagger}\phi+\phi_t\phi^{\dagger})\\\nonumber
&-\lambda^2\left(16\phi\phi^{\dagger}\mathcal{V}-4[\phi_x^{\dagger2}\phi^2+\phi_x^2\phi^{\dagger2}+2\phi_x\phi_x^{\dagger}\phi\phi^{\dagger}
+2i\phi\phi^{\dagger}(\phi_t^{\dagger}\phi-\phi_t\phi^{\dagger})] \right)\\\nonumber
&+16i(\phi_x\phi^{\dagger}-\phi\phi_x^{\dagger})+16.
\end{align}
We can check explicitly that the perturbative expansion of (\ref{eq:closedformJE}) match the results from the direct approach.

\section{Dynamical coordinates and gauge fields}
\label{sec:DCC}
In the relativistic case, $\rT\overline{\rT}$ deformation can be seen as a dynamical or field-dependent change of coordinates \cite{Dubovsky:2017cnj,Conti:2018tca,Guica:2019nzm}. However, such an interpretation is only valid on-shell (for a more detailed discussion, see \cite{Caputa:2020lpa}), it has several important applications. At the classical level, the dynamical change of coordinates gives yet another way to derive the classical deformed Lagrangian \cite{Coleman:2019dvf} as well as finding solutions to the deformed equation of motion and analyze the deformed classical symmetries \cite{Guica:2020uhm}. At the quantum level, it can be used to derive the deformed S-matrix \cite{Dubovsky:2017cnj}, at least in flat spacetime. In this section, we will show that $\rT\overline{\rT}$ deformation of non-relativistic QFTs also has such an interpretation. For the other two bilinear deformations, which involve the current $\mathcal{J}^{\mu}$, the interpretation is also interesting. In addition to the change of coordinates, one also needs to make a dynamical change of the $\mathrm{U}(1)$ gauge field $m_{\mu}$. More explicitly, in flat spacetime, the undeformed gauge field is a pure gauge $m_{\mu}=\partial_{\mu}\theta$. Under the bilinear deformations, we have $\theta\mapsto\Theta$ where $\Theta$ is field dependent. Such interpretations first appear in the $J\bar{T}$ and $JT_a$ deformations of relativistic QFTs \cite{Guica:2017lia,Anous:2019osb}\footnote{Here we mean the undeformed theory is relativistic. The deformed theory is, of course, no longer Lorentz invariant.}. In what follows, we will first give the proposals of the dynamical change of coordinates and gauge fields and then apply them to find the deformed Lagrangians, which match our previous results. Finally, we apply them to find the deformed quantum S-matrix.\par

\subsection{The dynamical coordinates}
There are different ways to find the dynamical coordinates and gauge fields \cite{Dubovsky:2017cnj,Conti:2018tca,Guica:2017lia,Cardy:2019qao,Caputa:2020lpa}, one of which is provided by the gravity formulation. 
\paragraph{The $\rT\overline{\rT}$ deformation} Let us first discuss $\rT\overline{\rT}$ deformation. From the gravity formulation, we derive the saddle-point equations for the auxiliary fields $\tilde{\tau}_{\mu}$, $\tilde{e}_{\mu}$ (\ref{eq:defCurrent}). In the previous section, we fix $\tau_{\mu}=\delta_{\mu}^t,e_{\mu}=\delta_{\mu}^x$ in the flat gauge and solve for $\tilde{\tau}_{\mu}$, $\tilde{e}_{\mu}$. Alternatively, we could fix $\tilde{\tau}_{\mu}=\delta_{\mu}^t$, $\tilde{e}_{\mu}=\delta_{\mu}^x$ in (\ref{eq:defCurrent}), which leads to the following equations
\begin{align}
\label{eq:preDCCTTbar}
\tau_{\mu}=\delta_{\mu}^t+\lambda\epsilon_{\mu\nu}\mathcal{P}_0^{\nu},\qquad e_{\mu}=\delta_{\mu}^x-\lambda\epsilon_{\mu\nu}\mathcal{E}_0^{\nu},
\end{align}
where $\mathcal{P}_0^{\nu}$ and $\mathcal{E}_0^{\nu}$ are the currents in flat space (\ref{eq:flatCurrent}). These equations are no longer saddle-point equations for $\tilde{\tau}_{\mu}$ and $\tilde{e}_{\mu}$. Instead, following the intuition from the relativistic case, we can interpret them as defining a change of coordinates from $(x^1,x^2)=(t,x)$ to $(X^1,X^2)=(T,X)$ by setting $\tau_{\mu}=\partial_{\mu}X^1$ and $e_{\mu}=\partial_{\mu}X^2$  in (\ref{eq:preDCCTTbar}). Therefore, the dynamical change of coordinates $(t,x)\mapsto (T,X)$ is defined by
\begin{align}
\label{eq:Jacobian}
\partial_{\mu}T=\delta_{\mu}^t+\lambda\epsilon_{\mu\nu}\mathcal{P}^{\nu}_0,\qquad
\partial_{\mu}X=\delta_{\mu}^x-\lambda\epsilon_{\mu\nu}\mathcal{E}^{\nu}_0.
\end{align}
One important comment is that, to interpret $\tau_{\mu}$ and $e_{\mu}$ as derivatives of the new coordinates, they need to satisfy the consistency condition $\partial_{\mu}\partial_{\nu}X^a=\partial_{\nu}\partial_{\mu}X^a$. This is guaranteed by the conservation of the currents. For example, we need to check that $\partial_{\mu}\partial_{\nu}T=\partial_{\nu}\partial_{\mu}T$. Using (\ref{eq:Jacobian}), we have
\begin{align}
\partial_{\mu}\partial_{\nu}T-
\partial_{\nu}\partial_{\mu}T=
\partial_{\mu}\epsilon_{\nu\alpha}\mathcal{P}_0^{\alpha}-\partial_{\nu}\epsilon_{\mu\alpha}\mathcal{P}_0^{\alpha}=-\epsilon_{\mu\nu}\partial_{\alpha}\mathcal{P}_0^{\alpha},
\end{align}
which is zero due to the conservation equation $\partial_{\alpha}\mathcal{P}_0^{\alpha}=0$. Notice that the conservation is valid only on-shell, namely when the fundamental fields satisfy equations of motion. This implies the dynamical change of coordinate interpretation is an on-shell statement.

\paragraph{The hard rod deformation} Now, we consider the hard rod deformation. Again we consider the saddle-point equation of the gravity action and take the auxiliary fields in the flat gauge $\tilde{m}_{\mu}=\partial_{\mu}\theta$ and $\tilde{e}_{\mu}=\delta_{\mu}^x$, which leads to
\begin{align}
\label{eq:predccHR}
m_{\mu}=\partial_{\mu}\theta+\lambda\,\epsilon_{\mu\nu}\mathcal{P}^{\nu}_0,\qquad e_{\mu}=\delta_{\mu}^x-\lambda\epsilon_{\mu\nu}\mathcal{J}^{\nu}_0,
\end{align}
where $\mathcal{P}^{\nu}_0$ and $\mathcal{J}^{\nu}_0$ are the flat space currents (\ref{eq:flatCurrent}) \emph{with the modification}
\begin{align}
\partial_\mu\phi\mapsto D_{\mu}\phi=\left(\partial_{\mu}+\frac{i}{2}\partial_{\mu}\theta\right)\phi,\qquad
\partial_\mu\phi^{\dagger}\mapsto D_{\mu}\phi^{\dagger}=\left(\partial_{\mu}-\frac{i}{2}\partial_{\mu}\theta\right)\phi^{\dagger}.
\end{align}
In the $\rT\overline{\rT}$ deformation, we can simply take $\theta=0$ by gauge fixing. For the hard rod and JE deformations, we keep the pure gauge term explicitly for later convenience. Now we need to make a proper interpretation of (\ref{eq:predccHR}). We propose that it defines a change of coordinate $(t,x)\mapsto (t,X)$ together with a change of gauge $\theta\mapsto\Theta$ as follows
\begin{align}
\label{eq:DCChardrod}
\partial_{\mu}\Theta=\partial_{\mu}\theta+\lambda\,\epsilon_{\mu\nu}\mathcal{P}_0^{\nu},\qquad  \partial_{\mu}X=\delta_{\mu}^x-\lambda\epsilon_{\mu\nu}\mathcal{J}_0^{\nu}.
\end{align}
Notice that here only the spatial coordinate is transformed; the temporal direction is left invariant. This is very natural since $e_{\mu}$ is related to the spatial direction. Similarly, to make such interpretations we need to check the consistency relations $\partial_{\mu}\partial_{\nu}\Theta=\partial_{\nu}\partial_{\mu}\Theta$ and $\partial_{\mu}\partial_{\nu}T=\partial_{\nu}\partial_{\mu}T$ which follow from the conservation equations of the currents.

\paragraph{The JE deformation} This case is similar to the hard rod deformation. The relevant equations are
\begin{align}
m_{\mu}=\partial_{\mu}\theta+\lambda\,\epsilon_{\mu\nu}\mathcal{E}^{\nu},\qquad \tau_{\mu}=\delta_{\mu}^t-\lambda\epsilon_{\mu\nu}\mathcal{J}^{\nu}.
\end{align}
We propose that this corresponds to the coordinate transformation $(t,x)\mapsto(T,x)$ together with the gauge transformation $\theta\mapsto\Theta$
\begin{align}
\partial_{\mu}\Theta=\partial_{\mu}\theta+\lambda\,\epsilon_{\mu\nu}\mathcal{E}^{\nu},\qquad  \partial_{\mu}T=\delta_{\mu}^t-\lambda\epsilon_{\mu\nu}\mathcal{J}^{\nu}. \label{JE1}
\end{align}

\subsection{Deformed Lagrangian III. Dynamical coordinates}
In this subsection, we derive the deformed Lagrangians using the dynamical coordinates and gauge fields as a non-trivial check of our proposals (\ref{eq:Jacobian}), (\ref{eq:DCChardrod}), (\ref{JE1}). The derivation for the $\rT\overline{\rT}$ deformation is similar to the relativistic case. The generalizations to the other two deformations are new results. The calculation involves two steps. In the first step, we find the quantities $\partial_{\mu}X^a$ and $\partial_{\mu}\Theta$ in terms of fundamental fields and their derivatives. In the second step, we perform a change of coordinates/gauge field of the original Lagrangian to the new coordinates/gauge field and plug in the quantities that we found in the first step. This leads to the deformed Lagrangian in the new coordinates/gauge field.

\paragraph{The $\rT\overline{\rT}$ deformation} We first discuss how to obtain the Jacobian $\partial_{\mu}X^a$. Notice that the right hand side of (\ref{eq:Jacobian}) is given in terms of fundamental fields and their derivatives (\ref{eq:flatCurrent}). We rewrite the derivatives of the fields by the chain rule $\partial_{\mu}\phi=\partial_{X^a}\phi\,\partial_{\mu}X^a$ and $\partial_{\mu}\phi^{\dagger}=\partial_{X^a}\phi^{\dagger}\,\partial_{\mu}X^a$. Then (\ref{eq:Jacobian}) becomes an equation for $\partial_{\mu}X^a$, which can be solved explicitly in terms of $\phi,\phi^{\dagger}$ and $\partial_{X^a}\phi,\partial_{X^a}\phi^{\dagger}$. The solution is considerably more complicated than the relativistic case and can be found in the ancillary notebook.\par

After obtaining the Jacobian $\partial_{\mu}X^a$, we plug it into
\begin{align}
\mathcal{L}_{\lambda}^{\rT\overline{\rT}}=\frac{1}{\det(\partial_{\mu}X^a)}\left(\mathcal{L}(T,X)
+\lambda\,\epsilon_{\mu\nu}\mathcal{E}^{\mu}_0\mathcal{P}^{\nu}_0\right).
\end{align}
The determinant $\det(\partial_{\mu}X^a)$ comes from the change of the integration measure $\rd^2x\mapsto\det(\partial_{\mu}X^a)\rd^2X$ in the action. The quantities in the bracket are written in terms of $\partial_{X^a}\phi$, $\partial_{X^a}\phi^{\dagger}$ by the chain rule. After plugging in the explicit forms of the Jacobian, we obtain the deformed Lagrangian \eqref{eq:closedTTbar} in the new coordinates.

\paragraph{The hard rod deformation} Let us first explain how to obtain the quantities $\partial_{\mu}\Theta$ and $\partial_{\mu}X$. Writing out the first equation of \eqref{eq:DCChardrod} explicitly
\begin{align}
\label{eq:solveTheta}
\partial_t\Theta=&\,\frac{i\lambda}{2}\left(D_t\phi\,\phi^{\dagger}
-\phi\,D_t\phi^{\dagger}\right)+\lambda\,D_x\phi D_x\phi^{\dagger}-\lambda\,\mathcal{V},\\\nonumber
\partial_x\Theta=&\,\frac{i\lambda}{2}\left(D_x\phi\,\phi^{\dagger}-\phi\,D_x\phi^{\dagger}\right),
\end{align}
where
\begin{align}
\label{eq:covariant}
D_{\mu}\phi=\partial_{\mu}\phi+\frac{i}{2}\partial_{\mu}\Theta\,\phi,\qquad
D_{\mu}\phi^{\dagger}=\partial_{\mu}\phi^{\dagger}-\frac{i}{2}\partial_{\mu}\Theta\,\phi^{\dagger}.
\end{align}
Plugging into (\ref{eq:solveTheta}), we can solve for $\partial_\mu\Theta$ in terms of the fundamental fields and their derivatives. The explicit results can be found in the ancillary file. To rewrite the second equation (\ref{eq:dccHR}) explicitly, we perform the change of coordinate $(t,x)\mapsto(t,X)$ where the new spacial coordinate $X$ depends on $t$ and $x$. We have the following chain rule
\begin{align}
\partial_t\phi(t,x)=\partial_t\phi(t,X)+\partial_x X\,\partial_X\phi(t,X),\qquad \partial_x\phi(t,x)=\partial_x X\,\partial_X\phi(t,X).
\end{align}
The second equation then can be written as
\begin{align}
\label{eq:dccHR}
\partial_x X=1+\frac{\lambda}{2}\phi\phi^{\dagger},\qquad 
\partial_tX=\frac{i\lambda}{2}\partial_x X\left(D_X\phi\phi^{\dagger}-\phi D_X\phi^{\dagger}\right),
\end{align}
where
\begin{align}
D_X\phi=\partial_X\phi+\frac{i}{2}\partial_X\Theta\,\phi,\qquad 
D_X\phi^{\dagger}=\partial_X\phi^{\dagger}-\frac{i}{2}\partial_X\Theta\,\phi^{\dagger}.
\end{align}
Here $\partial_X\Theta=\partial_x\Theta/\partial_xX$. We can solve (\ref{eq:dccHR}) explicitly, which leads to
\begin{align}
\partial_tX=\frac{i\lambda}{2}\left(\partial_X\phi\,\phi^{\dagger}-\phi\,\partial_X\phi^{\dagger}\right),\qquad \partial_xX=1+\frac{\lambda}{2}\phi\phi^{\dagger}.
\end{align}
After obtaining $\partial_{\mu}\Theta$, $\partial_{\mu}X$ in terms of $\phi,\partial_t\phi,\partial_X\phi$ and their conjugates, we plug into
\begin{align}
\frac{1}{\partial_x X}\left(\mathcal{L}(\Theta,X)+\lambda\epsilon_{\mu\nu}\mathcal{J}^{\mu}_0\mathcal{P}^{\nu}_0\right),
\end{align}
where $\partial_x X$ comes from the change of integration measure $\der t \der x \mapsto  \der t \der X$. In the bracket, we replace the partial derivative by the covariant counterparts (\ref{eq:covariant}) and perform the change of coordinate from $(t,x)$ to $(t,X)$. Going through these steps, we find the deformed Lagrangian (\ref{eq:closedFHR}).

\paragraph{The JE deformation} This case is similar to the hard rod deformation, and thus we will be brief.
The chain rule for the change of coordinate $(t,x)\mapsto(T,x)$ is now
\begin{align}
\partial_t\phi(t,x)=\partial_t T\,\partial_T\phi(T,x),\qquad \partial_x\phi(t,x)=\partial_x\phi(T,x)+\partial_xT\,\partial_T\phi(T,x).
\end{align}
After finding the solution of $\partial_{\mu}\Theta$, $\partial_{\mu}T$, we plug into
\begin{align}
\frac{1}{\partial_t T}\left(\mathcal{L}_0(\Theta,T)+\lambda\epsilon_{\mu\nu}\mathcal{J}^{\mu}_0\mathcal{E}^{\nu}_0\right).
\end{align}
This reproduces (\ref{eq:closedformJE}).

\subsection{Deformed quantum S-matrices}
As another application for the dynamical coordinates/gauge field, we derive the deformed S-matrix in this section. The main steps parallel the derivations for the relativistic case \cite{Dubovsky:2017cnj,Anous:2019osb}, adapted to the non-relativistic settings. The deformed S-matrix for non-relativistic QFTs have been derived using other methods, see \cite{Jiang:2020nnb,Cardy:2020olv}. It is shown that the effect of solvable bilinear deformations on the S-matrix is by multiplying a CDD like phase factor. We will confirm these results from the dynamical coordinate point of view.

\paragraph{$\rT\overline{\rT}$ deformation}
To define the S-matrix, we need the notion of asymptotic states. Let us consider the asymptotic in-states. In the far past $t\to-\infty$, the fields are free and allow the following mode expansion
\begin{equation}
\label{eq:expansion1}
\phi_{{\rm in}}=\int\frac{\rd p}{\sqrt{4\pi\omega_p}} a_{{\rm in}}(p)\,e^{-it\omega_p+ixp-im\theta},\qquad
\phi_{{\rm in}}^{\dagger}=\int\frac{\rd p}{\sqrt{4\pi\omega_p}} a_{{\rm in}}^{\dagger}(p)\,e^{it\omega_p-ixp+im\theta},
\end{equation}
where $\omega_p=p^2$. Several remarks are in order. Firstly, notice that the mode expansion of each field involves only one type of ladder operators instead of both. This is due to the non-relativistic nature of the QFTs under consideration. Secondly, we included a background gauge potential $\theta$ in the mode expansion, which can in principle be absorbed as a normalization of the field. Here we put it explicitly because, under the hard rod and JE deformations, we have the additional transformation $\theta\mapsto\Theta$, which has non-trivial effects. It is, therefore, convenient to include them in the first place. The Fourier modes in the expansion of $\phi_{\text{in}}$ satisfies the free Schr\"odinger equation
\begin{align}
i D_t\phi=-D_x^2\phi,\qquad D_{\mu}\phi=(\partial_\mu+im\partial_{\mu}\theta)\phi,
\end{align}
where $m$ can be seen as the mass of the particle. In the previous sections, we have put $m=1/2$, here it is more convenient to leave it generic.\par

Under $\rT\overline{\rT}$ deformation, we are equivalently putting the theory on the new coordinates $(T,X)$. Therefore it is more natural to perform the mode expansion in terms of the new coordinates
\begin{equation}
\label{eq:expansion2}
\phi_{{\rm in}}=\int\frac{\rd p}{\sqrt{4\pi\omega_p}} A_{{\rm in}}(p)\,e^{-i\omega_pT+ipX+im\theta},\qquad
\phi_{{\rm in}}^{\dagger}=\int\frac{\rd p}{\sqrt{4\pi\omega_p}} A^{\dagger}_{{\rm in}}(p)\,e^{i\omega_pT-ipX-im\theta}.
\end{equation}
Notice that for the $\rT\overline{\rT}$ deformation, the gauge field is left untouched.

Comparing the two expansions (\ref{eq:expansion1}) and (\ref{eq:expansion2}), we find that the two sets of modes are related by
\begin{equation}
    A_{{\rm in}}^{\dagger}(p)=a_{{\rm in}}^{\dagger}(p)\,e^{i\omega_p \Delta T-ip\Delta X},
\end{equation}
where $\Delta T=t-T$ and $\Delta X=x-X$. From the definition of the dynamical coordinates (\ref{eq:Jacobian}), we have
\begin{align}
\partial_{\mu}(\Delta T) =-\lambda\epsilon_{\mu\nu}\mathcal{P}^{\nu}_0,\qquad
\partial_{\mu}(\Delta X) =+\lambda\epsilon_{\mu\nu}\mathcal{E}^{\nu}_0.
\end{align}
In the far past $t\to-\infty$ we can integrate along the spacial direction and obtain
\begin{align}
\label{eq:DeltaTX}
\Delta T(x)={\rm const}_1 +\lambda\int_{-\infty}^{x}\mathcal{P}^{0}(x'){\rm d}x',\\\nonumber
\Delta X(x)={\rm const}_{2}-\lambda\int_{-\infty}^{x}\mathcal{E}^{0}(x'){\rm d}x'.
\end{align}
We introduce the notations
\begin{align}
P_<(x)=\int_{-\infty}^x\mathcal{P}^0(x')\rd x',\qquad P_>(x)=\int_x^{\infty}\mathcal{P}^0(x')\rd x',
\end{align}
where $P<(x)$ and $P_>(x)$ measure the total momentum to the left and right of $x$. We define $E_<(x)$ and $E_>(x)$ similarly. The integration constants in (\ref{eq:DeltaTX}) can be chosen arbitrarily. For convenience, we follow \cite{Dubovsky:2017cnj,Anous:2019osb} and choose the constants in a parity symmetric way
\begin{align}
{\rm const}_{1} =-\frac{\lambda}{2}\int_{-\infty}^{\infty}\mathcal{P}^{0}(x)\,{\rm d}x,\qquad
{\rm const}_{2} =+\frac{\lambda}{2}\int_{-\infty}^{\infty}\mathcal{E}^{0}(x)\,{\rm d}x.
\end{align}
We then have
\begin{align}
\Delta T(x)=-\frac{\lambda}{2}\left(P_{>}(x)-P_{<}(x)\right),\qquad
\Delta X(x)=-\frac{\lambda}{2}(E_{<}(x)-E_{>}(x)).
\end{align}
In the infinite past, the spatial ordering of the particles is equivalent to their momentum ordering, which is special to 1+1 dimensional physics.
Keeping this in mind, we have \footnote{There is a relative sign between the two terms in the exponent since in our definition of $\mathcal{E}^\mu$, the eigenvalue of $\mathcal{E}^0$ is $ - \sum_i p_{i}^{2}$.}
\begin{equation}
A_{{\rm in}}^{\dagger}(p_k)=a_{{\rm in}}^{\dagger}(p_{k})
\times\exp\left[-\frac{i\lambda}{2}\left(\sum_{j=1}^{k-1}(e_j p_k-p_j e_k)+\sum_{j=k+1}^{N}(e_k p_j - e_j p_k)\right)\right ],
\end{equation}
where $e_k=\omega_{p_k}=p_k^2$ is the energy of the $k$-th particle.
As a result, the deformed and undeformed in-states are related by a phase factor
\begin{align}
|\{p_j\}_{\text{in}}\rangle_{\lambda}=
\exp\left(-i\lambda\sum_{j<k}(e_jp_k-p_je_k)\right)|\{p_j\}_\text{in}\rangle_0,
\end{align}
which takes the same form as in the relativistic case.
A similar analysis can be done for the out-states. Therefore, we conclude under $\rT\overline{\rT}$ deformation, the S-matrix is deformed in the same way as in relativistic QFT
\begin{align}  
    \mathbb{S}_{{\lambda} }^{\rT\overline{\rT}}\left(\{p_{i}\},\{{\bar{p}}_{j}\}\right)=
    e^{-i\lambda\sum_{j<k}(e_j p_k-p_j e_k)} e^{-i\lambda\sum_{j<k}(\bar{e}_j\bar{p}_k-\bar{p}_j\bar{e}_k)}
\,\mathbb{S}_{0}\left(\{p_{i}\},\{\bar{p}_{j}\}\right), \label{momentum}
\end{align} 
where $\{p_i\}$ and $\{\bar{p}_j\}$ are the momenta for the in- and out-states.

\paragraph{The hard rod deformation}
Now we consider the hard rod deformation. The mode expansion in the dynamical coordinates and gauge field is given by
\begin{align}
\label{eq:modeexpHR}
\phi_{\text{in}}=\int\frac{\rd p}{\sqrt{4\pi\omega_p}}A_{\text{in}}(p)e^{-i\omega_pt+ipX+im\Theta},\qquad
\phi_{\text{in}}^{\dagger}=\int\frac{\rd p}{\sqrt{4\pi\omega_p}}A^{\dagger}_{\text{in}}(p)e^{i\omega_pt-ipX-im\Theta}.
\end{align}
Notice that the time coordinate is unchanged in this case, but the gauge $\theta$ is changed. The two sets of modes are thus related by
\begin{align}
A_{\text{in}}^{\dagger}(p)=a_{\text{in}}^{\dagger}(p)e^{-ip\Delta X-im\Delta\Theta},
\end{align}
where $\Delta\Theta=\theta-\Theta$. From (\ref{eq:DCChardrod}), we find
\begin{align}
\label{eq:DeltaquanHR}
\partial_{\mu}(\Delta X) =+\lambda\epsilon_{\mu\nu}\mathcal{J}^{\nu},\qquad
\partial_{\mu}(\Delta\Theta) =-\lambda\epsilon_{\mu\nu}\mathcal{P}^{\nu}.
\end{align}
Integrating these equations along the spacial direction in the asymptotic past as before, we obtain
\begin{align}
\Delta X(x)=-\frac{\lambda}{2}(M_{<}(x)-M_{>}(x)),\qquad
\Delta\Theta(x)=-\frac{\lambda}{2}(P_{>}(x)-P_{<}(x)),
\end{align}
where $M_<(x)$ and $M_>(x)$ are the total mass to the left and right of $x$, respectively. Similar considerations like before lead to the following deformed S-matrix
\begin{align}
\mathbb{S}_{\lambda}^{\text{HR}}\left(\{p_i\},\{\bar{p}_j\}\right)=
e^{+i\lambda\sum_{j<k}(m_j p_k-p_j m_k)}e^{+i\lambda\sum_{j<k}(\bar{m}_j \bar{p}_k-\bar{p}_j \bar{m}_k)}\,\mathbb{S}_0\left(\{p_i\},\{\bar{p}_j\}\right),
\end{align}
where $\{m_k\}$ and $\{\bar{m}_k\}$ are the masses of the particles of the in- and out-states. For simple theories with only one type of particle, we have $m_k=\bar{m}_k=m$.\par

Let us comment on an important difference between the hard rod and $\rT\overline{\rT}$ deformation. In the hard rod case, the temporal direction is not modified. Therefore the non-locality only occurs in the spatial direction. Alternatively, we can integrate the first equation of (\ref{eq:DeltaquanHR}) by taking the integration constant to be zero. This leads to
\begin{align}
X(x)=x-\lambda\int_{-\infty}^{x}\mathcal{J}^0(x')\rd x'.
\end{align}
Let us assume for simplicity that there is only one type of particle with mass $m$ and take $\lambda>0$. In this case, we have
\begin{align}
X(x)=x-(\lambda m)\times\{\text{number of particles to the left of }x\}.
\end{align}
As alluded in the introduction, this new coordinate has an intuitive physical interpretation \cite{Jiang:2020nnb,Cardy:2020olv}. Suppose instead of considering a collection of point particles that we consider hard rods of size $m\lambda$. The new coordinate $X(x)$ is measuring the free space between the rods to the left of $x$. The phase factor in the deformed S-matrix precisely takes into account the fact that the `particle' now has a finite size. Therefore, the hard rod deformation makes the point particles to finite-sized hard rods, which is the origin of its name. For $\lambda<0$, the interpretation is that the distance between the particles is increased.\par

For $\rT\overline{\rT}$ deformation, similar interpretation applies to the change of coordinates in the spacial direction $x\mapsto X$. Namely, particles become hard rods under the deformation. The size of each rod is proportional to the energy of the rod. On the other hand, for $\rT\overline{\rT}$ deformation, the temporal coordinate is also transformed $t\mapsto T$. Therefore the non-locality is also in the time direction. For JE deformation, only the temporal coordinate is transformed. The physical interpretation for the change of coordinates in the temporal direction seems to be more subtle. This is also reflected in the fact that the deformed Lagrangians of the $\rT\overline{\rT}$ and JE deformations are much more complicated than the hard rod deformed one.

\paragraph{The JE deformation}
Finally, we consider the JE deformation. The mode expansion in the new coordinates and background gauge field is
\begin{align}
\phi_{\text{in}}=\int\frac{\rd p}{\sqrt{4\pi\omega_p}}A_{\text{in}}e^{-i\omega_pT+ipx+im\Theta},\qquad
\phi_{\text{in}}^{\dagger}=\int\frac{\rd p}{\sqrt{4\pi\omega_p}}A^{\dagger}_{\text{in}}e^{i\omega_pT-ipx-im\Theta}.
\end{align}
The two sets of modes are related by
\begin{align}
A_{\text{in}}^{\dagger}(p)=a_{\text{in}}(p)\,e^{i\omega_p\Delta T-im\Delta\Theta}.
\end{align}
Using (\ref{JE1}), we have
\begin{align}
\partial_{\mu}(\Delta T) =\lambda\epsilon_{\mu\nu}\mathcal{J}^{\nu},\qquad
\partial_{\mu}(\Delta\Theta) & =-\lambda\epsilon_{\mu\nu}\mathcal{E}^{\nu}.
\end{align}
Integrating along the spacial direction and choosing the parity symmetric integration constants, we find in the asymptotic past
\begin{align}
\Delta T(x)=-\frac{\lambda}{2}(M_{<x}-M_{>x}),\qquad
\Delta\Theta(x)=-\frac{\lambda}{2}(E_{>x}-E_{<x}).
\end{align}
It follows that the deformed S-matrix is given by
\begin{align}
\mathbb{S}^{{\rm {\rm JE}}}_{\lambda}\left(\{p_{i}\},\{\bar{p}_{j}\}\right)=e^{-i\lambda\sum_{j<k}(m_j e_k-m_k e_j)}
e^{-i\lambda\sum_{j<k}(\bar{m}_j \bar{e}_k-\bar{m}_k\bar{e}_j)}
\, \mathbb{S}_{0}\left(\{p_{i}\},\{\bar{p}_{j}\}\right).
\end{align}
We found that the deformed S-matrices for the three bilinear deformations have similar structures. In fact, this structure holds in more general situations. For integrable theories, one can construct similar solvable bilinear deformations using the higher conserved currents. They lead to similar CDD factors, as has been shown in \cite{Jiang:2020nnb}.

\section{Conclusions and discussions}
\label{sec:concl}
We have shown that for non-relativistic QFTs, three fundamental solvable bilinear deformations can be defined: $\rT\overline{\rT}$, hard rod, and JE. These deformations can be interpreted as coupling the undeformed QFT to specific Newton--Cartan geometries. The gravity formulations offer us a geometrical perspective for such deformations and provide us with powerful tools to compute important physical quantities. Classically, we computed the deformed Lagrangians in closed forms for the Schr\"odinger model with a generic potential. Quantum mechanically, we derived the deformed quantum S-matrices for the three deformations.\par

There are many future directions one can pursue based on the current work. One immediate question is the quantization of the Newton--Cartan gravities in this paper. This question is of great conceptual importance and may shed light on the relativistic case as well.
Technically quantizing Newton--Cartan gravity can be anticipated to be more tractable because of a preferred time foliation \cite{Gallegos:2019icg,Gomis:2019zyu,Grumiller:2020elf}.
Furthermore, it gives a new route to reach relativistic quantum gravity by considering relativistic corrections after quantization \cite{Kuchar:1980tw,Duval:1990hj}.
One exciting and concrete question in the non-relativistic context is the hard rod deformation of the free boson. On the one hand, it is shown that quantum mechanically, the deformed theory describes a collection of free hard rods \cite{Jiang:2020nnb,Cardy:2020olv}, interacting only when they touch each other. This quantum mechanical model has been known for a few decades and has been studied extensively (see for example \cite{Nagamiya,Sutherland,Wadati}). On the other hand, our work gives a rather different, although not completely unexpected, formulation of the model --- coupling the free boson to a non-relativistic gravity theory. This paper has mainly focused on the classical aspects of the gravity formulation, although we also derived the deformed S-matrix. It is interesting to study the quantum aspects of the gravity theory and make more direct contacts to the quantum hard rod model.\par

Another interesting direction is to explore the relations with $\rT\overline{\rT}$ deformations of relativistic theories. It is well-known that a non-relativistic QFT in $D$ dimension can be obtained from relativistic QFTs in at least two ways:
The first one is by a null reduction of a $D+1$-dimensional QFT, which is discussed in more detail in Appendix \ref{sec:null_reductions}; the other is by performing a $1/c$ expansion of a $D$-dimensional relativistic QFT where $c$ is the speed of light. In the context of solvable deformations, both relations are interesting to explore.\par

From the null reduction perspective, the Schr\"odinger model we considered in this paper can be obtained by a 3d Klein--Gordon model with a generic potential. The three fundamental currents $\mathcal{J}^{\mu},\mathcal{P}^{\mu}$ and $\mathcal{E}^{\mu}$ are different components of the stress-energy tensor of the 3d theory. It would be fascinating to see whether we can `uplift' the deformations we defined in this paper to the 3d theory in some proper sense. This might give a concrete clue for defining a $\rT\overline{\rT}$ like deformation for QFTs in higher dimensions, at least in 3d.\par

From the $1/c$ expansion perspective, we can start with a relativistic QFT, $\rT\overline{\rT}$ deform it, and perform the $1/c$ expansion. If such a procedure is well-defined, it gives another way to define a $\rT\overline{\rT}$ deformed theory. As an interesting example, it is known that the Lieb--Liniger model can be obtained from the Sinh--Gordon theory by taking the non-relativistic limit \cite{Kormos:2009eqa,Kormos:2009yp}. The $\rT\overline{\rT}$ deformation of the Sinh--Gordon model has been studied in \cite{Cavaglia:2016oda,Conti:2019dxg}. Therefore, it is interesting to take the $1/c$ expansion of the deformed Sinh--Gordon model and compare it with what we obtained in the current work. We expect the results to be rather different since it is known that the $1/c$ expansion is typically related to the Type II Newton--Cartan gravity, which is briefly studied in Appendix \ref{sec:typeII_NC}. It would be interesting to clarify the details and generalize to other theories.\par

Yet another interesting direction is to study the holographic dual of the deformed non-relativistic theories. In the relativistic case, there have been several proposals for the holographic dictionary \cite{McGough:2016lol,Kraus:2018xrn,Guica:2019nzm,Hirano:2020nwq}. It would be interesting to see how these proposals are generalized to the non-relativistic cases. We emphasize that the hard rod and JE deformations are new in the non-relativistic contexts, and it would be exciting to see their holographic interpretations.
To explore the holographic dictionary, it is necessary first to study the deformed theories coupled to non-relativistic conformal Type I Newton--Cartan geometry, which has local Schr\"odinger symmetry \cite{Son:2008ye,Hartong:2014pma,Hartong:2014oma,Bergshoeff:2014uea}.
Such theories exhibit a larger symmetry and should be under more analytical control. As a result, more physical quantities can be computed. It would be interesting to compute physical quantities such as the spectrum \cite{Cavaglia:2016oda,Smirnov:2016lqw}, partition functions \cite{Datta:2018thy,Aharony:2018bad,Aharony:2018ics,Hashimoto:2019hqo,Hashimoto:2019wct,Caputa:2019pam}, and correlation functions \cite{Aharony:2018vux,Guica:2019vnb,Cardy:2019qao,He:2019ahx,He:2020udl,Li:2020pwa,Kruthoff:2020hsi,Hirano:2020ppu} more explicitly in these cases.

Finally, it is known that the $\rT\overline{\rT}$ deformation has a deep connection with string theory, see for example \cite{Giveon:2017nie,Baggio:2018gct,Apolo:2018qpq,Chakraborty:2018vja,Apolo:2019zai,Sfondrini:2019smd,Callebaut:2019omt}. It would be interesting to explore similar connections between our three bilinear deformations and the non-relativistic string theories \cite{Andringa:2012uz, Bagchi:2013bga,Harmark:2017rpg,Bergshoeff:2018yvt,Blair:2019qwi,Gallegos:2019icg,Gomis:2019zyu,Roychowdhury:2020kma}. Some of these questions have been investigated recently in \cite{Blair:2020ops} for the $\rT\overline{\rT}$ deformation. We believe our results in this paper will be helpful to pursue this direction further.\par

\paragraph{Note added} While finishing this paper, we became aware of the upcoming paper \cite{Roberto:toAppear}, which has some partial overlap with our work.
In particular, the dynamical change of coordinates and the deformed classical Lagrangian of the $\rT\overline{\rT}$ deformation are derived independently.

\subsection*{Acknowledgements}
YJ would like to thank Yang Zhang for initiating the collaboration with JX.
The work of DH is supported by the Swiss National Science Foundation through the NCCR SwissMAP.
We would like to thank Paolo Ceschin, Riccardo Conti and especially Roberto Tateo for helpful correspondences.

\appendix
\section{Review of \texorpdfstring{$D$}{D}-dimensional Newton--Cartan geometry}\label{sec:Newton-Cartan_app}
In this Appendix, we review Newton--Cartan geometry in general $D=d+1$ dimensions, of which there are actually two related but distinct types.
We will study the significance of torsion and non-metricity in the connections, where the situation is fundamentally different from Lorentzian geometry.
Finally, we will study matter theories on these geometries, their currents, and on-shell Ward identities.

\subsection{Newton--Cartan geometries}\label{sec:NC_geometries_app}
Newton--Cartan geometry essentially consists of the metrics\footnote{In section \ref{sec:NC_2d} we considered $e^\mu$ as the fundamental object instead of $h^{\mu\nu}$, but this was only because in $D=2$ we have that since $h^{\mu\nu}$ is of rank 1, it factorises as $h^{\mu\nu}=e^\mu e^\nu$.}
$\tau_{\mu},\,h^{\mu\nu}$ satisfying 
\begin{equation}
    \tau_{\mu}h^{\mu\nu}=0
\end{equation}
and their projective inverses $-v^{\mu},\,h_{\mu\nu}$ \cite{Cartan1,Cartan2}.
$\tau_{\mu}$ is called the clock-form and gives the local direction of time.
The total time elapsed for an observer following her world-line $\gamma$ between two events $A$ and $B$ is
\begin{equation}\label{sec:int_tau_observer}
    \Delta t_{AB} = \int_\gamma \tau = \int_{t_A}^{t_B} \tau_\mu\left({x(t)}\right) \dot x^\mu (t) \der t,
\end{equation}
which need not be the same as that of another observer with worldline $\gamma^\prime$ between the same events because of local time-dilation.
On the other hand, $h^{\mu\nu}$ is the inverse spatial metric, which is degenerate and of corank 1 because of $\tau_{\mu}h^{\mu\nu}=0$.

The metrics satisfy the completeness relations
\begin{equation}
-\tau_{\nu}v^{\mu}+h^{\mu\rho}h_{\rho\nu}=\delta_{\nu}^{\mu},
\end{equation}
and they transform as
\begin{align}
\delta\tau_\mu & = \mathcal{L}_\xi\tau_\mu,\label{eq:trafo_NC1}\\
\delta h_{\mu\nu} & = \mathcal{L}_\xi h_{\mu\nu}+\tau_\mu\lambda_\nu+\tau_\nu\lambda_\mu,\label{eq:trafo_NC2}\\
\delta v^{\mu} & = \mathcal{L}_{\xi}v^{\mu}+ h^{\mu\nu}\lambda_\nu,\\
\delta h^{\mu\nu} & = \mathcal{L}_\xi h^{\mu\nu},\label{eq:trafo_NC4}
\end{align}
where $\lambda_\mu$ satisfying $v^\mu \lambda_\mu=0$ is the Galilean boost parameter and $\xi^\mu$ is a diffeomorphism generating vector field as in \eqref{eq:Lie_derivative_diffeo}.
$\tau_{\mu},\,h^{\mu\nu}$ are thus tensorial, while $v^{\mu},\,h_{\mu\nu}$ transform under Galilean boosts, corresponding to ambiguity in defining them as proper inverses.

We can define the (inverse) vielbeine with Galilean frame bundle covariance as
\begin{align}
E_{\mu}^{A} & = \left(\tau_{\mu},\,e_{\mu}^{a}\right),\qquad
E_{A}^{\mu}  = \left(-v^{\mu},\,e_{a}^{\mu}\right),
\end{align}
where the spatial vielbeine now gets a spatial frame index $a,b,\ldots =\{1,\ldots,d=D-1\}$.
Written out in components the completeness relations are
\begin{align} \label{eq:NC_comleteness_rel_general}
    &\tau_\mu e^\mu_b = 0, \qquad
    \tau_\mu v^\mu = -1, \qquad
    e^a_\mu e_b^\mu = \delta^a_b, \qquad
    e^a_\mu v^\mu = 0, \qquad
    -\tau_\mu v^\nu + e^a_\mu e_a^\nu = \delta_\mu ^\nu.
\end{align}
The spatial metrics are related to the vielbeine as
 \begin{equation}
     h^{\mu\nu} := \delta^{ab} e^\mu_a e^\nu_b,\qquad
     h_{\mu\nu} := \delta_{ab} e_\mu^a e_\nu^b.
 \end{equation}
Flat spatial indices can be raised and lowered at will with $\delta^{ab}$ and $\delta_{ab}$, but the same is not true for the zero indices.
The vielbeine transforms under local Galilean transformations as
\begin{align}
\delta\tau_{\mu} & = \mathcal{L}_{\xi}\tau_{\mu},\\
\delta e_{\mu}^{a} & = \mathcal{L}_{\xi}e_{\mu}^{a}+\lambda^{a}{}_b e_{\mu}^{b}+\lambda^{a}\tau_{\mu},\\
\delta v^{\mu} & = \mathcal{L}_{\xi}v^{\mu}+e_{a}^{\mu}\lambda^{a},\\
\delta e_{b}^{\mu} & = \mathcal{L}_{\xi}e_{b}^{\mu}+\lambda_{b}{}^a e_{a}^{\mu},
\end{align}
where $\lambda^{ab}$ is a local spatial rotation and $\lambda^{a} = e^{\mu a} \lambda_\mu$ is a local Galilean boost.
We review the Galilei algebra and other non-relativistic algebras in Appendix \ref{sec:NR_groups}.
The elemental Newton--Cartan geometry and its curvature tensors can also elegantly be obtained as the gauging of the Galilei algebra \cite{Hartong:2015zia}.

Depending on what extra fields there are present, we can define two distinct types of Newton--Cartan geometry, whose properties we review below\footnote{There is also many other interesting extended geometries, see for example \cite{Bergshoeff:2018vfn,Bergshoeff:2019ctr}.}.

\subsubsection{Type I}\label{sec:typeI_NC}
Type I Newton--Cartan geometry is the type we have considered in the main text, where the local frame bundle symmetry is the Bargmann group \cite{Havas:1964zza,Duval:1984cj,DePietri:1994je,Andringa:2010it}.
We have here that the extra field is the $\mathrm{U}(1)$ gauge field $m_{\mu}$.
It is closely related to the Newtonian potential with $\Phi=-v^{\mu}m_{\mu}$ being the Newtonian potential.
It transforms as
\begin{equation}\label{eq:m_trafo}
    \delta m_{\mu} = \mathcal{L}_{\xi}m_{\mu}+\lambda_{a}e_{\mu}^{a}+\partial_{\mu}\sigma ,
\end{equation}
where $\sigma$ is an arbitrary function.

The geometry can likewise be obtained by gauging the Bargmann algebra \cite{Andringa:2010it}.
It can also conveniently be embedded in a $D+1$ dimensional Lorentzian spacetime with a null Killing vector as we will review in Appendix \ref{sec:null_reductions}.
It is \emph{not} the geometry that describes Newtonian gravity in a covariant formulation unless $\mathrm{d}\tau=0$, but it has other applications as advertised in the main text.
See \cite{Hansen:2018ofj,Hansen:2020pqs} for more details.

\subsubsection{Type II}\label{sec:typeII_NC}
Type II Newton--Cartan geometry is the geometry that arises in the large speed of light $c$ expansion of general relativity when expanded in $1/c^2$ in a covariant fashion \cite{Dautcourt:1996pm,Tichy:2011te,VandenBleeken:2017rij,Hansen:2018ofj}.
In addition to the fundamental Newton--Cartan metrics $\tau_\mu,\,v^\mu,\,h_{\mu\nu},\,h^{\mu\nu}$, which are leading order, there are also two next-to-leading order gauge fields $m_{\mu}$ and $\bar{\Phi}_{\mu\nu}$.
The $m_{\mu}$ here is \emph{not} a $\mathrm{U}(1)$ gauge field as it has additional terms proportional to $\der \tau$ compared to \eqref{eq:m_trafo}; it is a so-called torsional-$\mathrm{U}(1)$ gauge field.
The transformation properties of $m_{\mu}$ only coincides with its Type I cousin when $\der \tau=0$.
The fields all arise from the large speed of light expansion\footnote{One can also do the more general $1/c$ expansion, which introduces several new fields, see \cite{VandenBleeken:2019gqa,Ergen:2020yop}.} in $1/c^{2}$ of \emph{any} $D$-dimensional Lorentzian metric, which can be expanded systematically as
\begin{align}
g_{\mu\nu} & =  -c^{2}\tau_{\mu}\tau_{\nu}+\bar{h}_{\mu\nu}+c^{-2}\bar{\Phi}_{\mu\nu}+\mathcal{O}\left(c^{-4}\right),\nonumber \\
g^{\mu\nu} & =  h^{\mu\nu}-c^{-2}\left(\hat{v}^{\mu}\hat{v}^{\nu}+h^{\mu\rho}h^{\nu\sigma}\bar{\Phi}_{\rho\sigma}\right)+\mathcal{O}\left(c^{-4}\right),
\end{align}
where\footnote{These fields can likewise be defined in Type I Newton--Cartan geometry, where they are also boost invariant, but not $\mathrm{U}(1)$ invariant.}
\begin{align}
\bar{h}_{\mu\nu} & :=  h_{\mu\nu}-2\tau_{(\mu}m_{\nu)},\label{eq:boost_inv_metrics1} \\
\hat{\Phi} & :=  -v^{\mu}m_{\mu}+\frac{1}{2}h^{\mu\nu}m_{\mu}m_{\nu},\\
\hat{v}^{\mu} & :=  v^{\mu}-h^{\mu\nu}m_{\nu},\label{eq:boost_inv_metrics3}
\end{align}
all are invariant under local Galilean boosts and satisfy the completeness relation \cite{Hartong:2015wxa,Hartong:2015zia}
\begin{equation}
     -\tau_{\nu}\hat{v}^{\mu}+h^{\mu\rho}\bar{h}_{\rho\nu}=\delta_{\nu}^{\mu}.
\end{equation}
With these, it is easy to form objects that are guaranteed to be Galilean boost invariant.

The Lorentzian metric can, in principle, be expanded to any order.
The gauge transformations of the next-to-leading order fields $m_{\mu}$ and $\bar{\Phi}_{\mu\nu}$ can be traced back to be the result of subleading diffeomorphisms.
This is the geometry that describes Newtonian gravity in a covariant formulation.
When $\mathrm{d}\tau=0$, $\bar{\Phi}_{\mu\nu}$ decouples and the two geometries coincides, but they are distinct with different couplings to matter.
Relativistic matter fields can be expanded in $1/c^2$ in a similar spirit.
One can then also expand the Lagrangian of the matter fields coupled to Lorentz geometry systematically order-by-order.
The higher the order, the more relativistic effects are taken into account.

From now on, we focus on Type I Newton--Cartan geometry, although many statements are identical or have closely related equivalents in Type II.
We refer the reader to for example \cite{VandenBleeken:2017rij,Hansen:2018ofj,Hansen:2019svu,Hansen:2019vqf,VandenBleeken:2019gqa,Hansen:2020pqs} for more information and applications.

\subsection{Connections and curvatures}\label{sec:general_connections_curvatures}
To form diffeomorphic invariant actions, we must introduce covariant derivatives~$\nabla_\mu$.
It is natural to require the covariant conservation of the metrics through
\begin{align}
    0&=\nabla_\rho \tau_\mu = \partial_\rho \tau_\mu - \Gamma^\lambda_{\rho\mu}\tau_\lambda,\label{eq:cov_consv_metricity1}\\
    0&=\nabla_\rho h^{\mu\nu} = \partial_\rho h^{\mu\nu} + \Gamma^\mu_{\rho\lambda}h^{\lambda\nu} + \Gamma^\nu_{\rho\lambda}h^{\mu\lambda},\label{eq:cov_consv_metricity2}
\end{align}
where $\Gamma^\rho_{\mu\nu}$ is the affine connection.
Any connection that satisfies this is a Newton--Cartan metric-compatible connection \cite{Bekaert:2014bwa,Bekaert:2015xua,Hansen:2020wqw}.
Notice that this does not imply that $v^\mu$ and $h_{\mu\nu}$ are covariantly constant because of the degenerate metric structure.
Equation \eqref{eq:cov_consv_metricity1} implies that the temporal part of torsion of a Newton--Cartan connection is \emph{always} fixed to be
\begin{equation}
    (\der \tau)_{\mu\nu} = 2\partial_{[\mu} \tau_{\nu]} = \tau_\rho \Gamma_{[\mu\nu]}^\rho.
\end{equation}
There is no equivalent of the Levi-Civita connection because there is no unique solution to requiring torsionlessness and metricity.
Contrary to Lorentzian connections, we can see from \eqref{eq:cov_consv_metricity1} that requiring a torsionless connection actually puts a constraint on the temporal vielbein.
This serves as another indication that both torsion and non-metricity are natural features of Newton--Cartan geometry, a statement that can be made precise by studying Galilean frame bundles \cite{Hansen:2020wqw}.
Depending on the properties of $(\der \tau)_{\mu\nu}$ we may subdivide Newton--Cartan geometry into three different classes:
\begin{description}
    \item[$\pmb{\der \tau =0}$:] This is torsionless Newton--Cartan geometry. In this class, there exists a notion of absolute time $t$ as we have (up to topological obstructions)
    \begin{equation}
        (\der \tau)_{\mu\nu} = 0 \qquad\implies\qquad    \tau_\mu = \partial_\mu t.
    \end{equation}
    As can be seen from \eqref{sec:int_tau_observer}, all observers agree on the time interval between two events independent of their worldline as the closedness of $\tau_\mu$ implies $\oint \tau =0$.
    \item[$\pmb{\tau \wedge\der \tau =0}$:] This is known as twistless torsional Newton--Cartan geometry (TTNC). In general, we have $\oint \tau \neq 0$, so observers experience local time dilation. However, there is a foliation of spacetime into spatial hypersurfaces of simultaneity as is guaranteed locally by the Frobenius theorem. We can, without loss of generality, write
    \begin{equation}
        \tau_\mu = \mathrm{e}^{-\Psi}\partial_\mu T,
    \end{equation}
    where $\Psi=\Psi(x)$ is known as the Luttinger potential measuring the local time dilation and $T=T(x)$ is the time-function, which can also be taken as a coordinate.
    \item[$\pmb{\tau\wedge\der \tau \neq 0}$:] This is known as general torsional Newton--Cartan geometry (TNC). It is acausal because locally any two points can be connected with space-like curves, i.e., one with tangent vectors $\tau_\mu \dot x^\mu=0$ \cite{Geracie:2015dea}. For our purposes of coupling a field theory to a background geometry, this is the relevant one: Only in this case can we do completely arbitrary variations as $\tau_\mu$ is unconstrained.
\end{description}

The canonical choice for a connection is
\begin{equation}
\check\Gamma_{\mu\nu}^{\lambda}:=-v^{\lambda}\partial_{\mu}\tau_{\nu}+\frac{1}{2}h^{\lambda\sigma}\left(\partial_{\mu}h_{\nu\sigma}+\partial_{\nu}h_{\mu\sigma}-\partial_{\sigma}h_{\mu\nu}\right),\label{eq:C_connection}
\end{equation}
which has both torsion
\begin{equation}
    2\check\Gamma_{[\mu\nu]}^{\lambda} = -2v^{\lambda}\partial_{[\mu}\tau_{\nu]}
\end{equation}
and non-metricity, as besides $\check\nabla_\mu \tau_\nu = \check\nabla_\mu h^{\nu\rho}=0$ we have
\begin{equation}
\check\nabla_\mu v^\nu=\frac{1}{2}h^{\nu\rho}\mathcal{L}_v h_{\rho\mu},\qquad\check\nabla_\mu h_{\nu\rho}=\tau_{(\nu}\mathcal{L}_v h_{\rho)\mu},
\end{equation}
where $\mathcal{L}_v$ is the Lie derivative along the flow of $v^\mu$.
The connection \eqref{eq:C_connection} is $\mathrm{U}(1)$ invariant as it is only built from the vielbeine, but it transforms under local Galilean transformations.
This is in contrast with the Levi-Civita connection, which is invariant under local Lorentzian transformations.
It is still, in a sense, the closest we get to a "Levi-Civita connection" in NC geometry:
It has the minimal torsion allowed as the spatial torsion $2e_\lambda^a\check\Gamma_{[\mu\nu]}^{\lambda}=0$ is zero.

Another natural connection to work with is the manifestly boost invariant connection
\begin{equation}
\bar{\Gamma}_{\mu\nu}^{\lambda}:=-\hat{v}^{\lambda}\partial_{\mu}\tau_{\nu}+\frac{1}{2}h^{\lambda\sigma}\left(\partial_{\mu}\bar h_{\nu\sigma}+\partial_{\nu}\bar h_{\mu\sigma}-\partial_{\sigma}\bar h_{\mu\nu}\right),\label{eq:bar_connection}
\end{equation}
where the boost invariant fields $\hat{v}^{\lambda},\,\bar h_{\mu\nu}$ are given by \eqref{eq:boost_inv_metrics1},\eqref{eq:boost_inv_metrics3}.
The connection is torsionful with
\begin{equation}
    2\bar\Gamma_{[\mu\nu]}^{\lambda} = -2\hat v^{\lambda}\partial_{[\mu}\tau_{\nu]}
\end{equation}
and non-metricity, as besides $\bar\nabla_\mu \tau_\nu =\bar\nabla_\mu h^{\nu\rho}=0$ we have
\begin{align}
\bar\nabla_\rho v^\mu&=\frac{1}{2}h^{\mu\lambda}\mathcal{L}_{\hat v} \bar h_{\rho\lambda} -h^{\mu\lambda}\left(\tau_{\rho}\partial_{\lambda}\hat{\Phi}-\hat{\Phi}(\der \tau)_{\rho\lambda}\right),\\
\bar\nabla_\rho \bar h_{\mu\nu} &=\tau_{(\mu}\mathcal{L}_{\hat v} \bar h_{\nu)\rho}
+2\hat{\Phi}(\der \tau)_{\rho(\mu}\tau_{\nu)}-2\tau_{\mu}\tau_{\nu}\partial_{\rho}\hat{\Phi}-2\tau_{\rho}\tau_{(\mu}\partial_{\nu)}\hat{\Phi}.
\end{align}
The connection depends on $m_\mu$ linearly through \eqref{eq:boost_inv_metrics1}-\eqref{eq:boost_inv_metrics3} and is thus not $\mathrm{U}(1)$ invariant: It is not possible to write down a form where both symmetries are manifest simultaneously using just Type I fields.
One thus needs to work harder to guarantee the Bargmann symmetry of field theories; see \cite{Hartong:2015wxa,Geracie:2016inm} for more comments.

Whatever choice of connection one makes, we can define a Riemann curvature the usual way through the commutator of the covariant derivatives \cite{Hansen:2020pqs}:
\begin{align}
\left[\nabla_\mu,\nabla_\nu\right]X_\sigma  = & R_{\mu\nu\sigma}{}^\rho X_\rho-2\Gamma^\rho{}_{[\mu\nu]}\nabla_\rho X_\sigma,\\
\left[\nabla_\mu,\nabla_\nu\right]X^\rho  = & -R_{\mu\nu\sigma}{}^\rho X^\sigma-2\Gamma^\sigma_{[\mu\nu]}\nabla_\sigma X^\rho,
\end{align}
where $X^\mu$ as a vector and $X_\mu$ a covector.
In components, this gives
\begin{equation}
R_{\mu\nu\sigma}{}^{\rho}:=-\partial_{\mu}{\Gamma}_{\nu\sigma}^{\rho}+\partial_{\nu}{\Gamma}_{\mu\sigma}^{\rho}-{\Gamma}_{\mu\lambda}^{\rho}{\Gamma}_{\nu\sigma}^{\lambda}+{\Gamma}_{\nu\lambda}^{\rho}{\Gamma}_{\mu\sigma}^{\lambda}.\label{eq:Riemann_tensor}
\end{equation}
The Ricci tensor can also always be defined as
\begin{equation}\label{eq:Ricci_tensor}
R_{\mu\nu} := R_{\mu\rho\nu}{}^{\rho},
\end{equation}
and as the Newton--Cartan connections are both torsionful and non-metric, the antisymmetric part is in general non-zero.

\subsection{Matter currents of field theories on Newton--Cartan backgrounds}\label{subsec:Type-I-NC_currents}
Different choices for what set of fields one varies exist and each choice defines a different set of currents as the response to said variations.
Using the boost invariant fields $\hat v^\mu,\,h^{\mu\nu},\, \hat\Phi$ one will get a set of currents with all indices down, with manifest boost invariance \cite{Geracie:2014nka,Jensen:2014aia, Bergshoeff:2016lwr, Hartong:2015wxa}.

The canonical set of currents, in the sense that they in flat gauge are the Noether currents and all spacetime indices are raised, is defined as the response to varying $\tau_{\mu},\,e_{\mu}^{a},\,m_{\mu}$:
\begin{equation}
\delta_{\mathrm{bgd}}S\left[\varphi,\tau,e,m\right]:=\int_{M}\mathrm{d}^{D}x\,e\left(\mathcal{E}^{\mu}\delta\tau_{\mu}+\mathcal{P}^{\mu}{}_a\delta e_{\mu}^{a}+\mathcal{J}^{\mu}\delta m_{\mu}\right),\label{eq:currents_typeI0}
\end{equation}
where $e:= \det(\tau_\mu,e^a_\mu)$ is the measure, or equivalently
\begin{equation} \label{eq:currents_typeI1}
\mathcal{E}^{\mu}  :=  e^{-1}\frac{\delta S}{\delta\tau_{\mu}}\, \spa
\mathcal{P}^{\mu}{}_a  :=  e^{-1}\frac{\delta S}{\delta e_{\mu}^{a}}\, \spa
\mathcal{J}^{\mu}  :=  e^{-1}\frac{\delta S}{\delta m_{\mu}},
\end{equation}
where here $\mathcal{E}^{\mu}$ is the energy current, $\mathcal{P}^{\mu}{}_a$
the momentum current and $\mathcal{J}^{\mu}$ the mass current. A
Galilean boost relates $\mathcal{P}^{\mu}{}_a$ and $\mathcal{J}^{\mu}$
through the on-shell Ward identity
\begin{equation}
\mathcal{P}^{\mu}{}_a\tau_{\mu}=-\mathcal{J}^{\mu}e_{\mu a},
\end{equation}
which shows that $\mathcal{P}^{\mu}{}_a$ should be thought of as the stress-mass current. In $D>2$, we also have a rotational on-shell Ward identity 
\begin{equation}
    0= e_{\mu [a} \mathcal{P}^{\mu}{}_{b]} ,
\end{equation}
telling us that the spatial components of the momentum current are symmetric, also well-known from studying the Noether currents in flat spacetime.

Diffeomorphism invariance of the Lagrangian guarantees that the currents
satisfies the following covariant conservation equations:
\begin{equation} \label{eq:consv_equation_2}
0 =  \partial_{\mu}\mathcal{E}^{\mu}+\left(e^{-1}\partial_{\mu}e\right)\mathcal{E}^{\mu},\qquad
0 =  \partial_{\mu}\mathcal{P}^{\mu}{}_a+\left(e^{-1}\partial_{\mu}e\right)\mathcal{P}^{\mu}{}_a,\qquad
0 = \partial_{\mu}\mathcal{J}^{\mu}+\left(e^{-1}\partial_{\mu}e\right)\mathcal{J}^{\mu},
\end{equation}
which can be written as a covariant Newton--Cartan derivative using for example the connections \eqref{eq:C_connection} or \eqref{eq:bar_connection} if one wants to make covariance explicit, using the formula
\begin{equation}
    e^{-1}\partial_\mu \left(e X^\mu \right) = \nabla_\mu X^\mu + 2 \Gamma_{[\mu \rho]}^\rho X^\mu,
\end{equation}
for any vector field $X^\mu$.

\section{Non-relativistic groups} \label{sec:NR_groups}
The hallmark of non-relativistic groups is that there is a notion of absolute time in the sense that there is no boost rescaling the time coordinate \cite{Geracie:2015xfa,Duval:1984cj}.
This is exactly \emph{not} the case for the Poincar\'e group $\group{ISO}{1,d}=\group{SO}{1,d}\ltimes \mathbb{R}^{1,d}$ because of its Lie algebra commutator
\begin{equation}
    \left[J_{0j},P_k\right] = -\delta_{jk}H,
\end{equation}
where $J_{0j}$ is a Lorentz boost, $P_k$ a spatial momentum and $H$ the Hamiltonian generator.

The simplest non-relativistic group is the $d$ dimensional Euclidean group $\group{ISO}{d}$ since there is no time direction.
Its Lie algebra generators consist of spatial momenta $P_i$ and spatial rotation generators $J_{ij}=-J_{ji}$.
The non-zero commutation relations are
\begin{align}
\left[J_{ij}, P_{k}\right]	&=	\delta_{ik}P_{j}-\delta_{jk}P_{i}, \\
\left[J_{ij},J_{kl}\right]	&= \delta_{ik}J_{jl}-\delta_{jk}J_{il}-\delta_{il}J_{jk}+\delta_{jl}J_{ik}.
\end{align}

When including the Hamiltonian $H$ and non-relativistic Galilean boosts generated by $G_i$, we get the Lie algebra of the Galilei group $\group{Gal}{d}=\left(\group{SO}{d}\ltimes \mathbb{R}^{d}\right)\ltimes \mathbb{R}^{1,d}$, which has some additional non-zero commutators given by
\begin{align}
\left[H,G_{i}\right]	&=	P_{i}, \\
\left[J_{ij},G_{k}\right]	&=	\delta_{ik}G_{j}-\delta_{jk}G_{i} .
\end{align}
The Galilei algebra can be obtained by \.In\"on\"u--Wigner contracting the Poincar\'e algebra~\cite{Inonu:1953sp}.

The Galilei group's central extension is non-trivial and of special interest and is called the Bargmann group $\group{Barg}{d}=\left(\group{SO}{d}\ltimes \mathbb{R}^{d}\right)\ltimes \left(\mathbb{R}^{1,d}\otimes \group{U}{1}\right)$ \cite{Bargmann:1954gh}.
Its Lie algebra has, in addition to the above non-zero commutation relations, a new non-zero one given by
\begin{equation}
    \left[P_{i},G_{j}\right]=N\delta_{ij},
\end{equation}
where $N$ is a central charge corresponding to particle number or mass.
The algebra can be obtained from a null reduction of the Poincar\'e algebra in $D+1$ dimensions as well, which lays the foundations for the next section.

There are also conformal extensions like the Galilean conformal algebra \cite{Duval:2009vt,Bagchi:2009my,Bagchi:2012xr} and the Schr\"odinger algebra, which has an anisotropic scaling \cite{Niederer:1972zz,Henkel:1993sg,Bergshoeff:2014uea}.
Both allow for infinite-dimensional Virasoro-like extensions \cite{Bagchi:2009ca}.

\section{Null reductions} \label{sec:null_reductions}
In this Appendix, we review how to define the null reduction of $D+1$ dimensional Lorentzian geometries with a null Killing vector to obtain a $D$ dimensional Type I Newton--Cartan geometry.
We also study how to derive Bargmann invariant matter theories from the null reduction of relativistic ones.

\subsection{Lorentzian spacetimes with a null Killing vector}
The Type I NC fields fit into the null reduction of $D+1$-dimensional Lorentzian metric
$g_{MN}$ with $M=\left(\mu,u\right)$ and $N=\left(\nu,u\right)$ \cite{Julia:1994bs,Balasubramanian:2010uk,Christensen:2013rfa}.
Any Lorentzian metric with a null isometry $\partial_{u}$ can be written in adapted coordinates as
\begin{eqnarray}
g_{MN} & = & \left(\begin{array}{cc}
g_{\mu\nu} & g_{\mu u}\\
g_{u\nu} & g_{uu}
\end{array}\right)=\left(\begin{array}{cc}
\bar{h}_{\mu\nu} & \tau_{\mu}\\
\tau_{\nu} & 0
\end{array}\right),\label{eq:metric_reduced_null}\\
g^{MN} & = & \left(\begin{array}{cc}
g^{\mu\nu} & g^{\mu u}\\
g^{u\nu} & g^{uu}
\end{array}\right)=\left(\begin{array}{cc}
h^{\mu\nu} & -\hat{v}^{\mu}\\
-\hat{v}^{\nu} & 2\hat{\Phi}
\end{array}\right),\label{eq:inverse_metric_reduced_null}
\end{eqnarray}
where the Galilean boost invariant objects $\bar{h}_{\mu\nu},\, \hat{v}^{\nu},\,\hat{\Phi}$ are defined in \eqref{eq:boost_inv_metrics1}-\eqref{eq:boost_inv_metrics3}.
Notice that we have the constraint $g_{uu}=0$, which means that we cannot do variations with respect to $g_{uu}$ without leaving the null hypersurface.
Thus, we can not obtain the current corresponding to $g_{uu}$ in the null reduced theory, but all remaining ones are available to us.

We can, of course, also decompose the Lorentzian metric in terms of vielbeine as
\begin{equation}
    g_{MN} = \eta_{\hat A \hat B} E_{M}^{\hat{A}} E_{N}^{\hat{B}},
\end{equation}
where the $D+1$-dimensional Lorentzian vielbeine with $\hat{A}=\left(A,u\right)=\left(0,a,u\right)$ and $\hat{B}=\left(B,u\right)=\left(0,b,u\right)$ can be written as
\begin{eqnarray}
E_{M}^{\hat{A}} & = & \left(\begin{array}{ccc}
\tau_{\mu} & e_{\mu}{}^a & -m_{\mu}\\
0 & 0 & 1
\end{array}\right),\nonumber \\
E_{\hat{A}}^{M} & = & \left(\begin{array}{ccc}
-v^{\mu} & e^{\mu}{}_{a} & 0\\
-m_{\mu}v^{\mu} & m_{\mu}e^{\mu}{}_a & 1
\end{array}\right),
\end{eqnarray}
where the inverse vielbein is solved such that the completeness relation holds.

These formulae are useful for null reducing relativistic theories on Lorentzian backgrounds to obtain non-relativistic theories on Type I NC backgrounds, as we shall now see an example of.

\subsection{Klein--Gordon theory}
We here want to obtain the Schr\"odinger model on a general Newton--Cartan background by performing a null reduction of the Klein--Gordon model \cite{Hartong:2014pma,Jensen:2014aia}.
Our stating point is the complex Klein--Gordon scalar field $\Psi\left(t,\boldsymbol{x},u\right)$ with potential $V\left(|\Psi|\right)$ coupled to Lorentzian geometry as
\begin{equation}
\hat{S}_{\mathrm{KG}}\left[\Psi,g\right]=\int\mathrm{d}^{D+1}x\sqrt{-g}\left(-g^{MN}\partial_{M}\Psi^{\dagger}\partial_{N}\Psi-V\left(|\Psi|\right)\right).\label{eq:action_KG}
\end{equation}
Since we want a Bargmann scalar of mass $m$, the higher-dimensional field must be taken to be of the form
\begin{equation}
\Psi\left(t,\boldsymbol{x},u\right)=e^{+imu}\phi\left(t,\boldsymbol{x}\right).
\end{equation}
Using the null reduction of the metric (\ref{eq:inverse_metric_reduced_null}), we can easily perform the null reduction decomposing the metric and taking derivatives of the scalar field.
The result is
\begin{equation}
S_{\mathrm{Schr}}=\int\mathrm{d}^{D}xe\,\biggl[imv^{\mu}\phi D_{\mu}\phi^{\dagger}-imv^{\nu}\phi^{\dagger}D_{\nu}\phi-h^{\mu\nu}D_{\mu}\phi^{\dagger}D_{\nu}\phi-V\left(|\phi|\right)\biggr],\label{eq:action_massive scalar}
\end{equation}
where we have defined the $\mathrm{U}(1)$-covariant derivative 
\begin{equation}
D_{\mu}\phi:=\partial_{\mu}\phi+imm_{\mu}\phi.
\end{equation}
Upon setting $m=1/2$ and $D=2$ one obtains the action \eqref{eq:schr_2d_action} considered in the main text.
In addition to local Galilean symmetry, we have $\mathrm{U}(1)$ symmetry under
\begin{eqnarray}
\phi\left(x\right) & \mapsto & e^{-im\sigma\left(x\right)}\phi\left(x\right) \\
m_{\mu} & \mapsto & m_{\mu}+\partial_{\mu}\sigma\left(x\right).
\end{eqnarray}

The relation to the relativistic Hilbert energy-momentum tensor found by varying the $D+1$-dimensional metric $g_{MN}$
\begin{equation}
    T_{\mathrm{Hil}}^{MN} := \frac{1}{2}\sqrt{-g}\frac{\delta S}{\delta g_{MN}},\label{eq:EH_EM_tensor}
\end{equation}
is
\begin{eqnarray}
\mathcal{E}^{\mu} & = & T_{\mathrm{Hil}}^{\mu u}-T_{\mathrm{Hil}}^{\mu\nu}m_{\nu}, \\
\mathcal{P}^{\mu}{}_a & = & T_{\mathrm{Hil}}^{\mu\nu}e_{\mu a},\\
\mathcal{J}^{\mu} & = & -T_{\mathrm{Hil}}^{\mu\nu}\tau_{\nu}.
\end{eqnarray}
This decomposition holds in general and can also be used to show how the lower-dimensional currents can be used to assembly the relativistic energy-momentum tensor.

\bibliographystyle{JHEP}
\bibliography{refNC}

\end{document}